\documentclass[10pt,pre,a4paper,twocolumn,superscriptaddress,showpacs]{revtex4-1}
\usepackage{amsmath}
\usepackage{amssymb}
\usepackage{amsthm}
\usepackage{mathtools}
\usepackage{graphicx}
\usepackage{grffile}%to compile figures with dots in fignames
\usepackage{color}
\usepackage[dvipsnames]{xcolor}

\begin{document}

\author{Claus Heussinger} \affiliation{Institute for theoretical
  physics, Georg August University Göttingen, Friedrich Hund Platz 1,
  37077 Göttingen, Germany} \title{Collapse of columns of granular
  rods}

\begin{abstract}
  Simulation results are presented on the collapse of granular columns
  composed of rod-like particles. Columns can be stable and free-standing
  if either the friction coefficient is large enough, or the rods long
  enough. Destabilizing gravitational forces are counteracted by
  increased frictional forces between the rods. Different to columns
  made of spherical particles this is possible, because rod contacts
  can slide along the rod axes to generate the necessary ``frictional
  cohesion''.
\end{abstract}

\maketitle

\section{Introduction}

The question about the stability of granular systems in the presence
of external forces is important in many different fields of
engineering and even every-day life. Examples are wall forces that
induce clogging in hoppers during silo
discharge~\cite{ashour17:_outfl}, dune formation in the desert because
of wind~\cite{kroy02:_minim_model_sand_dunes} or avalanches and
landslides as a consequence of gravitational forces.

On the continuum scale the key concept for stability against
gravitational forces is the angle of repose. It characterizes the
maximal angle with the horizontal, at which a free surface of granular
material is still stable. The value of this angle is not unique and
depends on the experimental conditions~\cite{al-hashemi18}. Two
extremes in this respect are the measurement of the angle of repose in
carefully prepared static granular piles just before yielding, or
after a large-scale yielding event has come to rest.

\begin{figure}[ht]
  \includegraphics[width=0.4\textwidth]{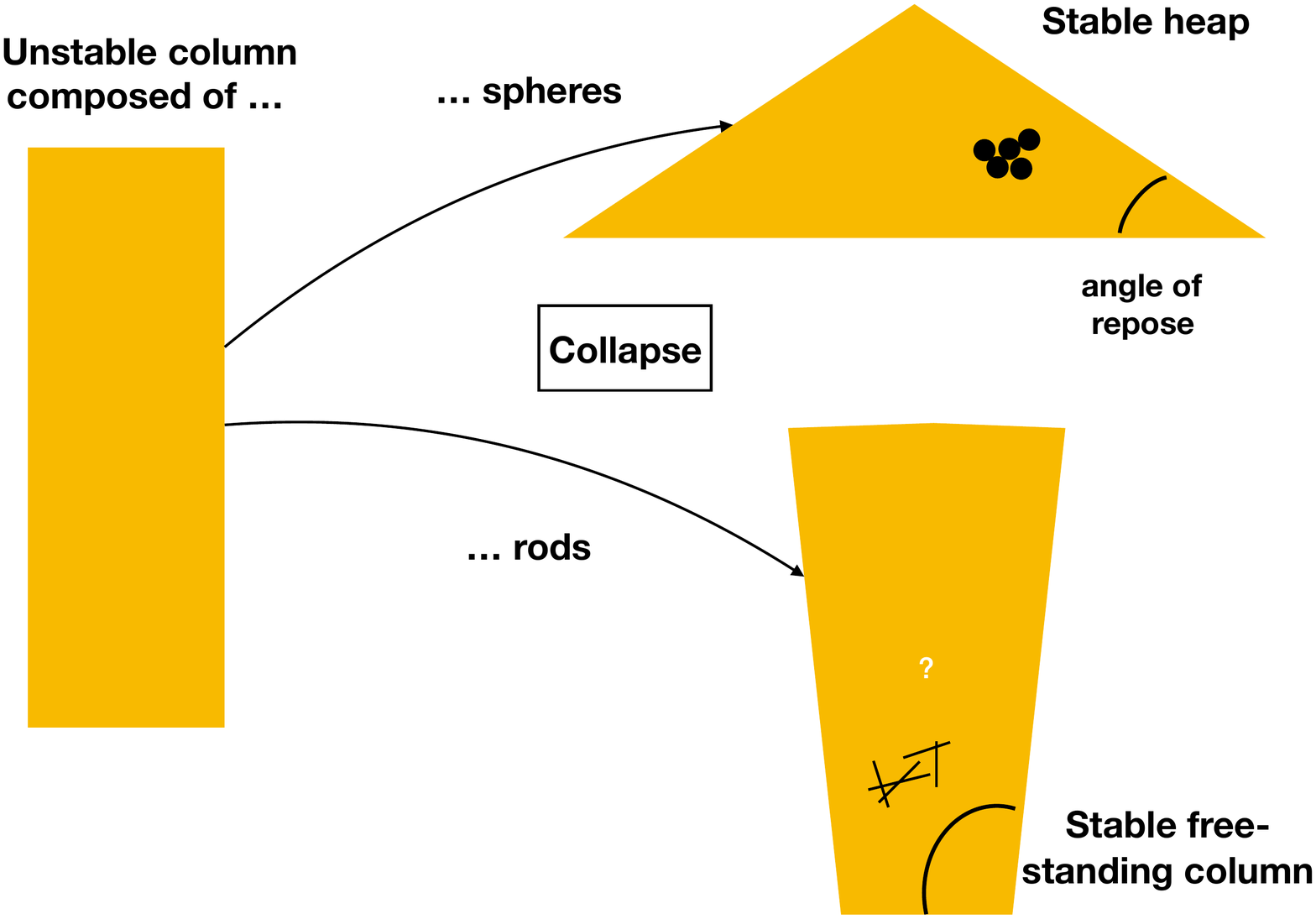}
  \caption{Sketch of the scenario of the simulations: a column of
    rod-like particles is generated and the container walls are
    removed. In the presence of gravity a column of spherical
    particles would collapse and form a flat heap. By way of contrast,
    a column of rods may be stable and form a free-standing
    quasi-solid structure.}\label{fig1}
\end{figure}

Here, we are interested in the latter problem (see Fig.\ref{fig1}). We
will generate a large column composed of granular particles. After
removing the confining walls free vertical surfaces are generated,
which are far above the critical angle of repose. As a consequence the
column will start to yield and collapse into a flat heap -- at least
for spherical or nearly-spherical grains. But what about grains that
take the shape of long thin rods?

The collapse of columns of (nearly-)spherical particles has been
studied with molecular dynamics simulations~\cite{lacaze08:_planar},
in continuum modeling~\cite{lagree11,holsapple13:_model} and many
experiments~\cite{xu16:_measur,lube07:_static,lajeunesse04:_spread}. Key
observable is the so-called run-out distance, i.e. the width of the
final collapsed heap.

When it comes to particles of different shapes, they are usually dealt
with in the context of jamming of random packings within containers or (in
simulations) periodic boundary
conditions~\cite{RevModPhys.82.2633,PhysRevE.97.012909,PhysRevE.75.051304,YUAN2019186,PhysRevMaterials.6.025603}.
%packings in containers
Stable packings of rods and their properties (density, contacts,
moduli, etc) have attracted a great deal of attention both in
experiments~\cite{Blouwolff_2006,parkhouse95,phi96,freeman19:_random}
and
simulation~\cite{willi03,PhysRevE.102.022903,PhysRevE.103.052903,zhao12:_dense,tangri2017packing}. Recent
work has discussed in detail the small-strain moduli of the assemblies
both without~\cite{PhysRevE.102.022903} and with frictional
inter-particle forces~\cite{PhysRevE.103.052903} and their dependence
on rod length $\ell$.

Non-convex particles assembled in the presence of gravity are known to
form stable free-standing columns, e.g.  u-shaped particles
\cite{PhysRevLett.108.208001} z-shaped particles
\cite{murphy16:_frees_z} or stars \cite{refId0}. These assemblies are
also discussed in the context of the design of new types of
load-bearing structures in architecture~\cite{dierichs21:_desig}.

Columns made of granular rods may also be stable and
free-standing. Experiments~\cite{desmond06:_jammin,trepanier10:_colum}
suggest a critical rod length above which columns are always stable
and do not collapse under the action of gravitational forces.
Therefore, rods are expected to undergo a transition to solid-like
behavior upon increasing rod length. Similar results have been
obtained when using granular chains as
particles~\cite{sarate22:_colum}. However, more detailed
investigations into what is happening at the collapse-to-solid
transition are missing. Simulations focus on the collapsed state
\cite{PhysRevE.85.061304,zhao18:_atten}, e.g. on the dependence of the
run-out distance on rod length.

\begin{figure*}[ht]
  \includegraphics[width=0.3\textwidth]{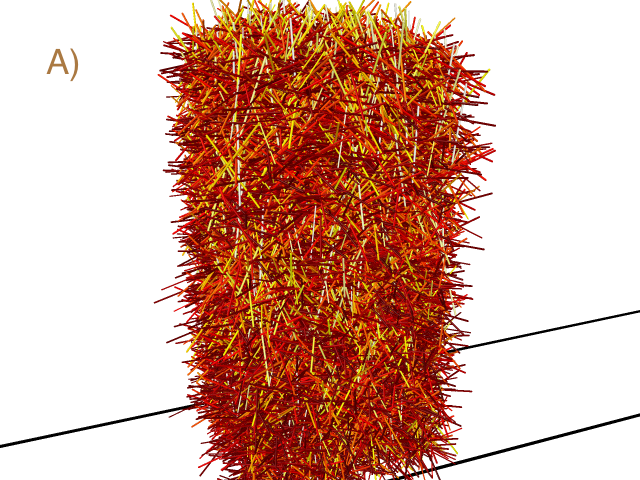}
  \includegraphics[width=0.3\textwidth]{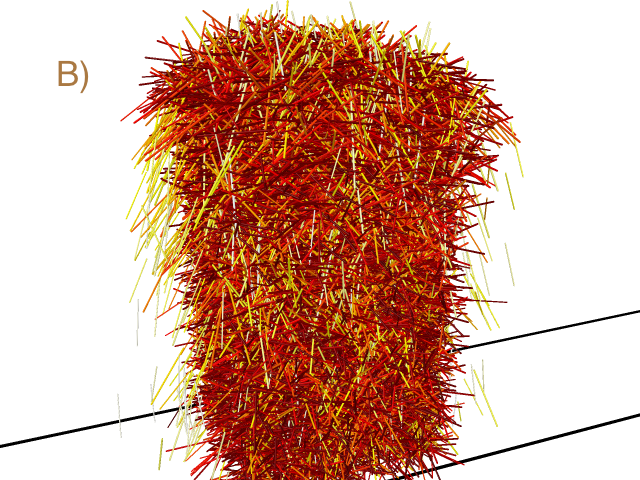}
  \includegraphics[width=0.3\textwidth]{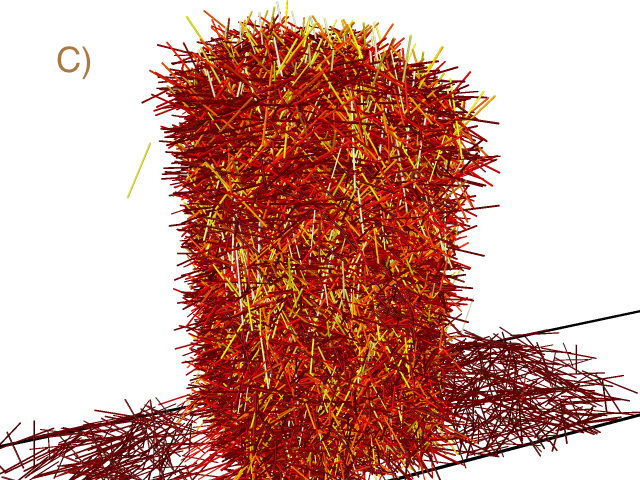}
  \caption{Yielding of a column after removal of the sidewalls
    (periodicity constraints in y-direction, in x-direction column
    remains periodic): (A) just after removal of the wall, the column
    compacts a bit; (B) intermediate stage, where unstable particles
    slide off the column (primarily from the top) one after the other;
    (C) final state; one of the last particles falls down. Rod length
    $\ell=40$, friction coefficient $\mu=0.1$. Particles are
    color-coded according to $\cos^2\theta$, where $\theta$ is the
    polar angle of the rod with the vertical direction. Simulation
    snapshots in this and the following figures are visualized with
    the software OVITO
    \cite{stukowski10:_visual_ovito_open_visual_tool}.}
  \label{fig:snapshots_stable}
\end{figure*}

\begin{figure*}[ht]
  \includegraphics[width=\textwidth]{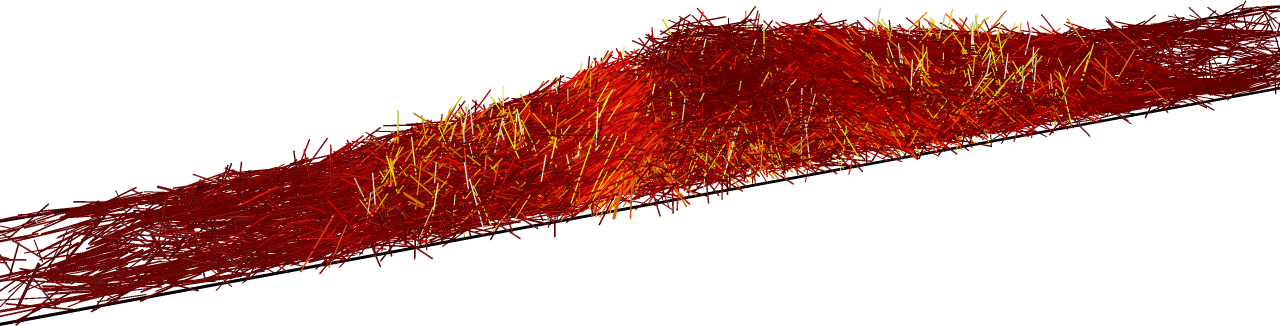}
  \caption{The same column as in Fig.~\ref{fig:snapshots_stable} but
    with a smaller friction coefficient $\mu=0.04$ collapses
    completely. The final state displays characteristic zones: in the
    centre particles are randomly oriented, reflecting the initial
    state; next to the center (orange zone) a deposit of particles
    that slid of the initial column is formed; in the wing several
    upright rods (yellow/white) are present that touch the ground;
    even further out, the heap ``runs out'' and all rods are nearly
    horizontal (dark red).}
  \label{fig:snapshots_unstable}
\end{figure*}

Here, we use computer simulations to investigate in detail the
granular column collapse for rod-like particles. The aim is to find
the relevant micro-structural observables and physical mechanisms that
lead to the observed transition from normal granular behavior -- full
collapse similar to spherical grains -- to the establishment of a
quasi-solid free-standing column with stable vertical free surfaces
(see Figs.~\ref{fig:snapshots_stable} and \ref{fig:snapshots_unstable}
for an illustration).

\section{Model}

A rod is modelled as a spherocylinder (SC), which consists of a
cylinder decorated by two hemi-spherical caps at the cylinder
ends. The center-line of the cylinder is called the backbone. Its
length $\ell$ determines the asphericity of the particle, ranging from
nearly spherical ($\ell\approx0$) to needle-like ($\ell\to\infty$). A
dimensionless measure is the aspect-ratio $\alpha=\ell/d$, where $d$
is the diameter of the cylinder.

\subsubsection{Interaction forces}

The interaction forces between two spherocylinders are calculated as
follows (details see \cite{PhysRevE.102.022903,PhysRevE.103.052903}): 

In a first step the shortest distance $\mathbf{r}$ between the
backbones of the two spherocylinders is determined. This determines
the site at which the interaction force acts. The force itself is
taken from models of spherical particles~\cite{cundall79}. The force
$\mathbf{f}_{ij}$ on particle $i$ from the contact with $j$ has
components normal $\mathbf{f}^n_{ij}$ and tangential
$\mathbf{f}^t_{ij}$ to the particle surface:
\begin{align}\label{eq:contactforce_definition}
  \mathbf{f}^n_{ij}&=[-k_n \delta_{ij} \mathbf{\hat e}_{ij} - c_n\mathbf{v}^n_{ij}] ,\\
  \mathbf{f}^t_{ij}&=[-k_t \boldsymbol{\xi} ^t_{ij} - c_t\mathbf{v}^t_{ij}].\nonumber
\end{align}
Here, the normal direction $\mathbf{\hat
  e}_{ij}=\mathbf{r}_{ij}/r_{ij}$ points from particle $j$ to $i$ at
the point of application of the force as determined by the shortest
distance. The normal overlap $\delta_{ij}= d_{ij}-r_{ij}$ is a
positive quantity. The tangential overlap $\boldsymbol{\xi} ^t_{ij}$
is the displacement tangential to the surface of the SC, which
accumulates during the lifetime of the contact. The parameters $k_n$
and $k_t$ are spring constants. A viscous damping force is present via
the parameters $c_n$ and $c_t$. It is proportional to the relative
velocity at the contact, which is split into normal
$\mathbf{v}^n_{ij}$ and tangential components $\mathbf{v}^t_{ij}$.

Frictional dissipation is taken into account by replacing
$\mathbf{f}^t$ by $\mu|\mathbf{f}^n|(\mathbf{f}^t / |\mathbf{f}^t|)$,
whenever the Coulomb inequality
\begin{equation}\label{eq:coulomb}
  |\mathbf{f}^t|<\mu |\mathbf{f}^n|\,,
\end{equation}
is violated.  The friction coefficient $\mu$ determines the strength
of frictional forces and is varied in the following.

\subsubsection{Rolling friction}

As described above frictional forces $\mathbf{f}^{t}$ only act when
the points of interaction on the two rods translate relative to each
other. Eq.~(\ref{eq:contactforce_definition}) does not apply when
interacting rods rotate relative to each other and around the point of
interaction (then $\boldsymbol{\xi} ^t$ and $\mathbf{v}^t$ in
Eq.~(\ref{eq:contactforce_definition}) vanish). In reality,
interaction sites are not points (like in the simulations) but have a
finite extent. This may give rise to torques that resist such
rotation, which is called rolling and twisting in this context. In
order to remove the kinetic energy in these degrees of freedom,
additional dissipative torques are incorporated in the
simulations. Without these effects particles would continue to roll or
twist forever, also across the surface.

In twisting motion the rotation vector is parallel to the surface
normal. A resistance builds up, because the contact zones on the
particles are rotating relative to each other. For rolling motion the
rotational component perpendicular to the surface normal enters. The
resistance to rolling is because of an asymmetry in the contact
pressure in the front and the back of the contact zone.

Many different models to capture these effects have been proposed~
\cite{PhysRevE.102.032903}. Here, we implement a simple viscous force
law that relates the twisting and rolling torques to the relative
angular velocities
$\boldsymbol{\omega}_{ij}=\boldsymbol{\omega}_i-\boldsymbol{\omega}_j$
of the two contacting particles.
\begin{eqnarray}\label{torque_roll_twist}
  \boldsymbol{\tau}_{ij}^t = -\gamma_t
  \boldsymbol{\omega}_{ij}^n &\qquad\qquad& \boldsymbol{\tau}_{ij}^r = -\gamma_r \boldsymbol{\omega}_{ij}^t
\end{eqnarray}

For the twisting $\tau_t$ and rolling torque $\tau_r$. The angular
velocities are split into normal (n) and tangential (t) contributions.
The physical dimension of the prefactors $\gamma_t$, $\gamma_r$ are
different from $c_n$, $c_t$ from the viscous force law from
translational motion. What is missing is of dimension length squared,
and is related to the size of the contact zone. We do not explicitly
model this dependence on contact zone size. The scope of our approach
is to inhibit indeterminate rolling or twisting motion by draining the
kinetic energy out of these degrees of freedom. For the special case
of $\gamma_t=\gamma_r\equiv\gamma$ Eq. (\ref{torque_roll_twist})
reduces to $ \boldsymbol{\tau}_{ij}^t+ \boldsymbol{\tau}_{ij}^r =
-\gamma(\boldsymbol\omega_i-\boldsymbol\omega_j)$ which is what we
adopt here because of computational efficiency.

\subsubsection{Particle-surface interactions}

Under the influence of gravity particles fall on a horizontal surface
and eventually come to rest. Particle-surface interactions are similar
to inter-particle interactions. Interaction sites with the surface are
the sperocylinder-end caps. Surface-tangential (xy-plane) and
surface-normal (z-direction) forces, as well as rolling friction are
defined as described above, with the only difference that the velocity
of the wall is zero. Rolling friction here inhibits, for example, that
rods roll indefinitely on the surface. The parameters used are
$\mu_W=10$, i.e. large friction limit, $c_t=c_n=2$
i.e. overdamped. Spring constants as well as rolling friction are as
between particles, $k_n=1$, $k_t=2k_n/7$, $\gamma=0.01$.

\subsection{Procedure}

The simulation proceeds in three steps. First, particles are
distributed randomly in space with volume fraction $\phi_0$. Overlaps
are removed as much as possible by running an energy minimization in
the absence of frictional, gravitational and surface forces (with
periodic boundary conditions in all three directions). Different
$\phi_0$ are used, most of the time somewhat smaller than the jamming
density of the particular assembly~\cite{PhysRevE.102.022903}. The
lateral system size is chosen such that $L_x=L_y > 3\ell$. System
height can be changed to generate columns of different aspect ratios
$L_z=aL_x$. Values used are $a=1,2,3$. Particle number $N$ is adjusted
accordingly and ranges from $N=6144\ldots18432$.

In the second step, frictional, gravitational and surface forces are
switched on and particles start to fall on a horizontal surface, which
is situated directly below the particle with the smallest
height. Periodic boundary conditions are retained in the horizontal (x
and y-directions) to mimic lateral walls. Particles are allowed to
settle and come to rest with these constraints. This represents the
initial state for the third step, the column collapse. Once settling
is complete, one of the periodicity constraints is removed (in y
direction), which generates two free vertical surfaces with normals in
+y and -y direction. This gives rise to additional yielding of the
column~\footnote{With the chosen boundary conditions the setting also
  resembles the dam-break problem, where collapse occurs in only one
  direction.}. The ensuing behavior ranges from complete collapse to
only slight compactification and overall stability of the free
surfaces. Figs.~\ref{fig:snapshots_stable} and
\ref{fig:snapshots_unstable} show examples of the yielding of a column
with $\ell=40$ and two different friction coefficients $\mu=0.1$
(Fig.~\ref{fig:snapshots_stable}) and $\mu=0.04$
(Fig.~\ref{fig:snapshots_unstable}), respectively. In the first case
the free surfaces are more or less stable with only a few particles
sliding off the structure. In the case of the smaller friction
coefficient the column collapses completely.

\section{Results}

\subsection{Transition from collapse to free-standing}

\begin{figure}[ht]
  \includegraphics[width=0.5\textwidth]{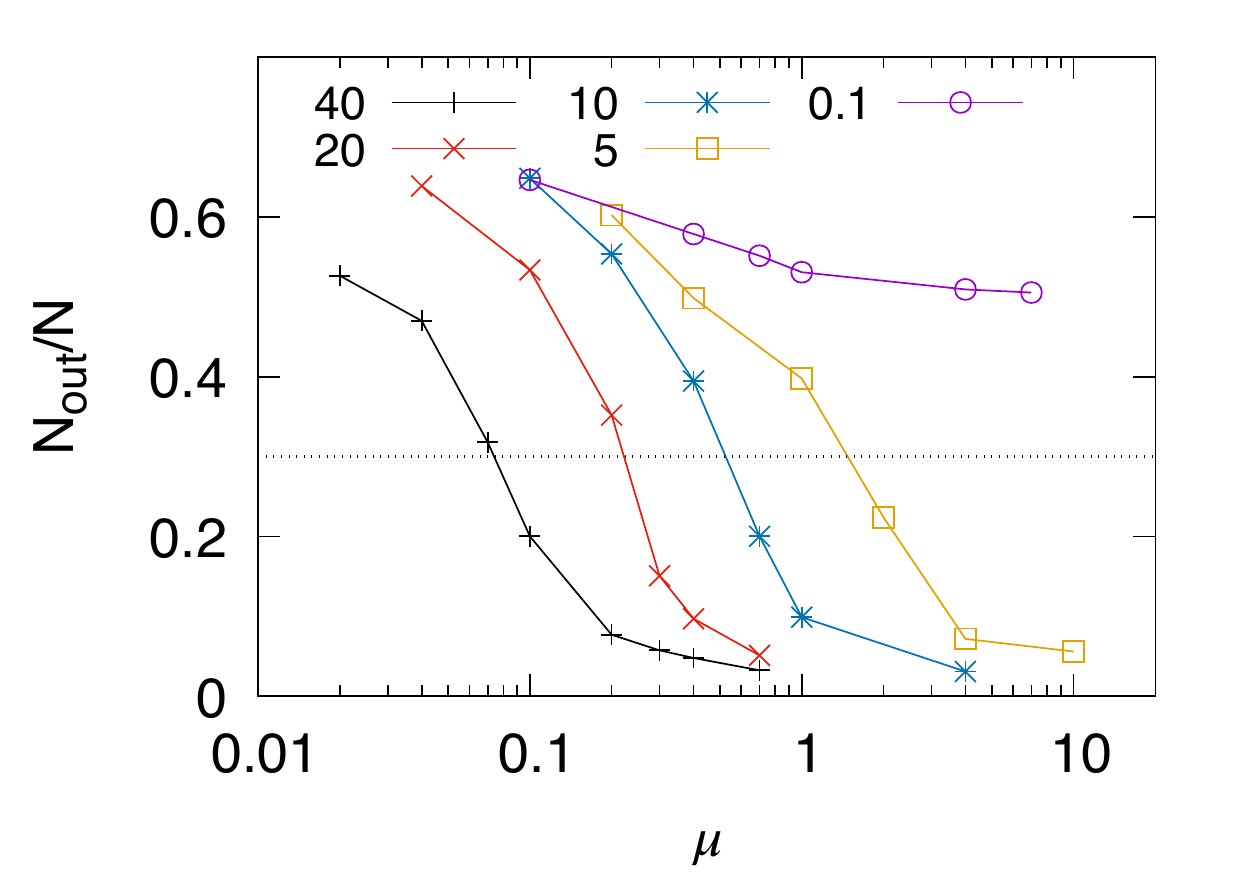}
  \caption{Average fraction $f=N_{out}/N$ of particles that leave the
    original simulation box, vs. $\mu$ for different $\ell=0.1 \ldots
    40$.}
  \label{fig:frac_mu_ell}
\end{figure}

We start with a discussion of columns with aspect ratio $a=1$,
i.e. cubic shape. A large set of different columns is prepared with
different starting density $\phi_0$, different random seed, length
$\ell$ and friction coefficient $\mu$. To quantify the (in-)stability
of the column the average fraction of particles $f=N_{\rm out}/N$ that
leave the initial simulation box is calculated (see
Fig.~\ref{fig:frac_mu_ell}). If no particles leave the initial box
($f=0$) the column is stable and unaffected by gravity. On the other
side, in a collapsing column many particles (but not all) would leave
the box.

By comparing columns made with different friction coefficient $\mu$ and
particle length $\ell$ one first key result appears: columns made of
long rods do not necessarily collapse in gravity. By decreasing the
friction coefficient a transition takes place from stable (or nearly
stable) columns to fully collapsed ones. Such a transition does not
occur for spheres or nearly-spherical particles (data-points with
$\ell=0.1$).

\begin{figure}[ht]
  \includegraphics[width=0.5\textwidth]{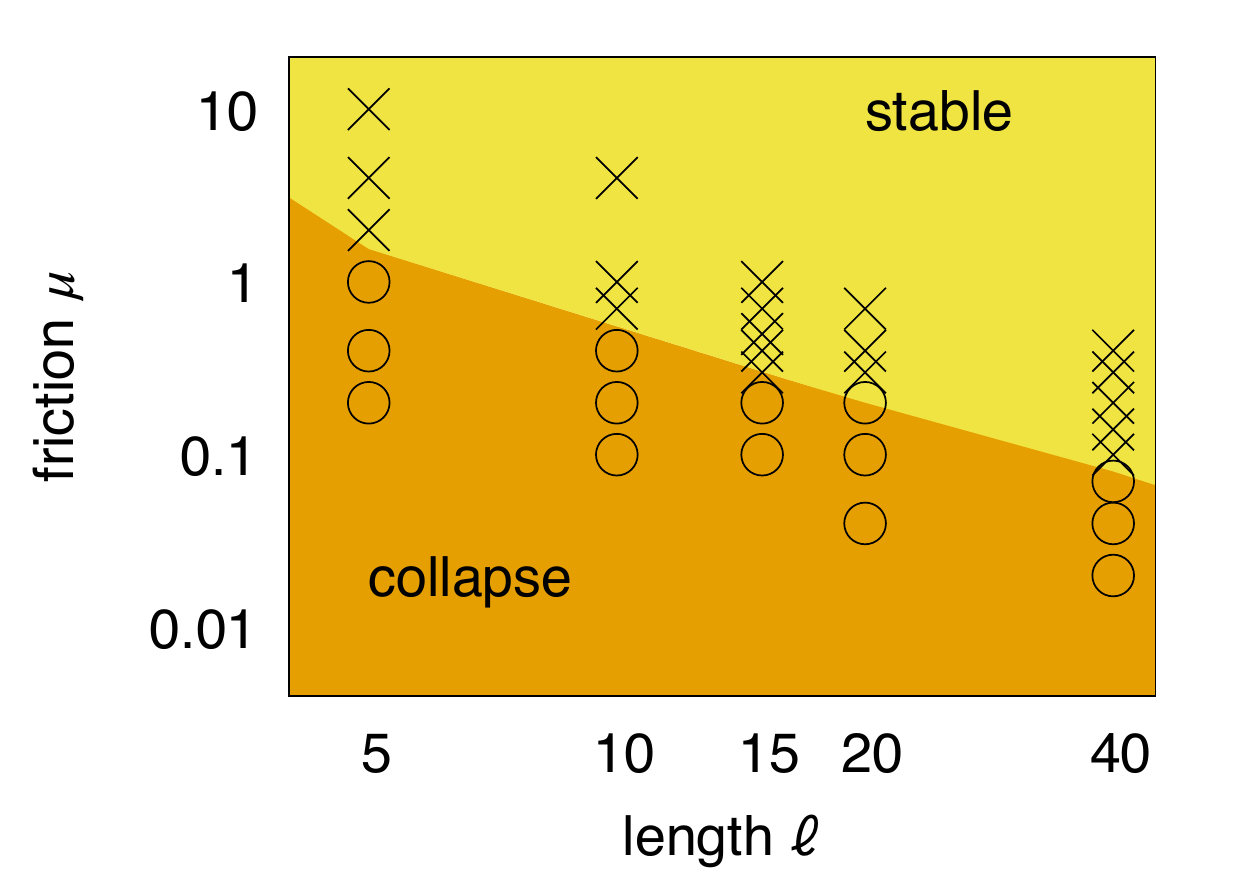}
  \caption{Stability phase diagram in the plane ($\ell,\mu$) as obtained
    from the analyis of Fig.~\ref{fig:frac_mu_ell}. The transition
    line $\mu_c(\ell)$ is approximate and just a guide to
    eye.}
  \label{fig:phase_mu_ell}
\end{figure}

By defining a (more or less arbitrary) threshold value for the
fraction $f$ (here 0.3, dotted horizontal line), a minimal friction
coefficient $\mu_c$ can be defined that is necessary to guarantee
stability of a column with given $\ell$. This way, two phases can be
distinguished in the plane $(\ell,\mu)$, separated by a transition
line $\mu_c(\ell)$ (Fig.~\ref{fig:phase_mu_ell}).

Two aspects of this phase diagram are immediately apparent. First,
increasing friction $\mu$ for a given rod length $\ell$ leads to more
stable columns. This trend seems quite natural, as higher $\mu$
implies higher frictional inter-particle forces that might contribute
to column stability. Still, such an effect is exclusive for long rods
and does \emph{not} occur in columns of (nearly-)spherical
particles. Second, the critical $\mu_c$ decreases with increasing
$\ell$, i.e.  columns with given friction are more stable when particles
are longer.

It is the goal of the remainder of this work to shed light on these
two trends. We will need to correlate column stability with fundamental,
local properties. Eventually, what is looked for are observables that
serve as predictors of collapse behavior. In other words, one would
want to know in how far the (in)stability after removal of the
sidewalls is already imprinted in the column when the confining walls
are still present.

\subsection{Gravity-settled state within container}

Therefore, let's start with a discussion of the gravity-settled state
within the container, i.e. after coming to rest but before removing
the side-walls. In the following the friction coefficient is fixed to
$\mu=0.1$ and rod length $\ell$ is varied from $\ell=10\ldots 40$. The
data is analyzed from columns with aspect ratio $a=3$, as in
Fig.~\ref{fig:snapshots_stable}. The value for $\mu$ is chosen, as
columns with $\ell=40$ (when removing the walls) remain stable while
$\ell=20$ collapse. For the latter, a higher friction $\mu=0.4$ is
needed to form a stable column.

Under the action of gravitational forces the initial column as a whole
falls onto the surface, compacts and comes to rest. Table
\ref{tbl:rel_compaction} compares initial and final densities in
columns with different $\ell$ and initial densities $\phi_0$. The
$\phi_0$ were chosen close to the (frictionless) jamming density of
the particular $\ell$. Overall, packings with longer rods display a
higher relative compaction. That is, long rods -- after coming to rest
-- live in a surrounding that is substantially denser than what they
started with. In addition, there is a dependence on the initial
density $\phi_0$. For given $\ell$, compaction is higher the smaller
the initial $\phi_0$.

\begin{table}[ht]
  \begin{tabular}{|c|c|c|c|}\hline
    \,\,length $\ell$\,\, & \,\,$\phi_0$ (initial)\,\, & \,\,$\phi_f$ (settled)\,\, &
    \,\,compaction\,\, \\\hline
    10 & 0.42 & 0.439 & 4.5\%\\
    10 & 0.41 & 0.431 & 5.1\%\\
    20 & 0.24 & 0.263 & 9.6\%\\
    20 & 0.23 & 0.256 & 11\%\\
    20 & 0.22 & 0.250 & 14\%\\
    40 & 0.1  & 0.126 & 26\%\\
    40 & 0.09 & 0.118 & 31\%\\
    40 & 0.08 & 0.112 & 40\%\\
    \hline
  \end{tabular}\caption{Volume fractions $\phi$ before and after
    settling in the container; $\mu=0.1$; relative compaction is
    largest in columns with smallest $\phi_0$. Longer rods compact
    stronger.}\label{tbl:rel_compaction}
\end{table}

\begin{figure}[ht]
  \includegraphics[width=0.5\textwidth]{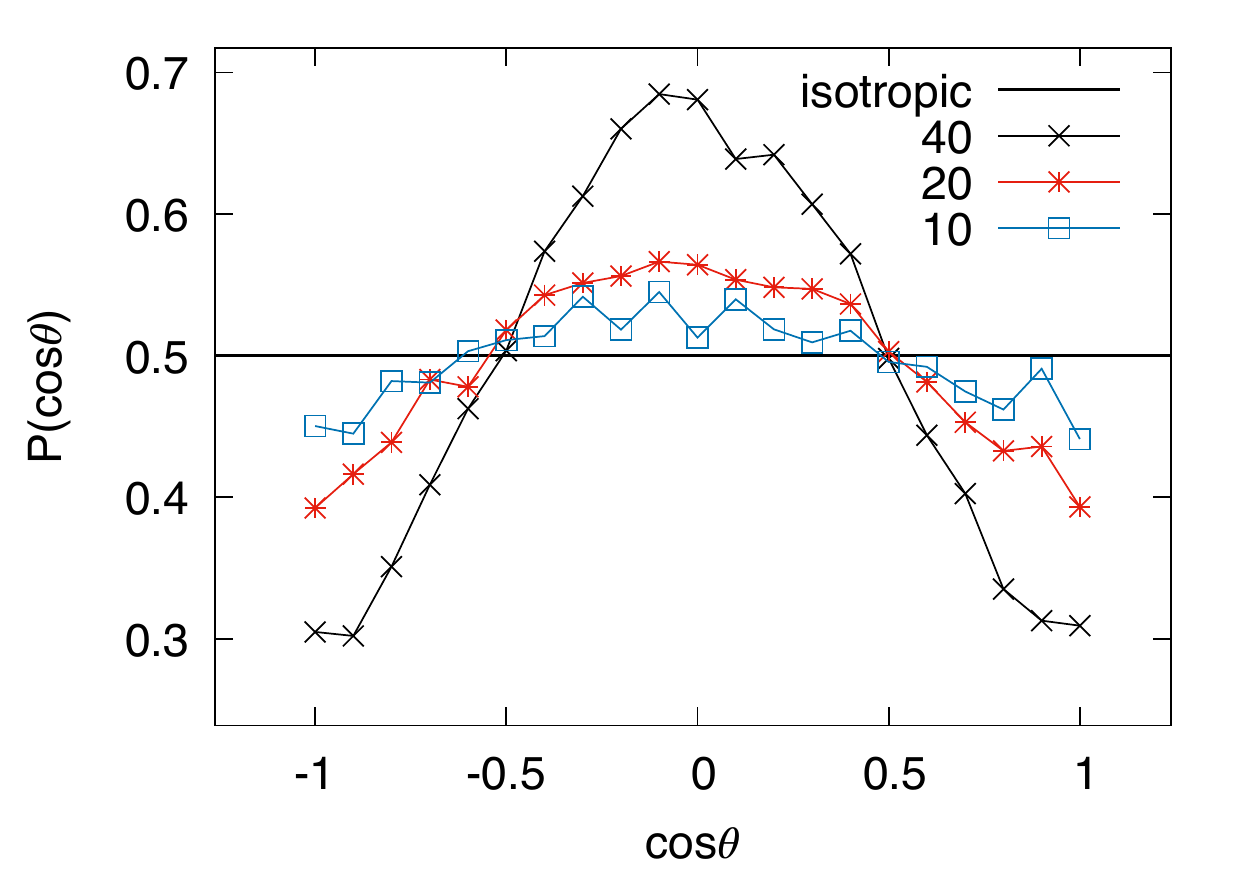}
  \caption{Probability distribution $P(\cos\theta)$ of rod orientation
    after settling within the container. The polar angle $\theta$ is
    measured with respect to the vertical.  Same systems as in
    Table~\ref{tbl:rel_compaction}. Data averaged over systems with
    different $\phi_0$.}
  \label{fig:orientation}
\end{figure}

One consequence of this variation in compaction is a difference in the
orientational distribution of the rods. While the starting state is
isotropic the final deposited state shows some anisotropy, with more
rods oriented towards the horizontal (Fig.~\ref{fig:orientation}).
A visual impression of the amount of ordering is given in
Fig.~\ref{fig:snapshots_stable}A. Horizontal rods are depicted in dark
red, while vertical in light yellow. To the eye, this system does not
seem to be highly anisotropic. A quantitative measure is given by the
order parameter $1-3\langle \cos^2\theta\rangle$, which is zero in the
isotropic state and 1 in the fully planar state, respectively
(negative values correspond to the usual nematic order). The system in
the figure has a value of 0.2 is therefore only moderately
anisotropic. All other systems, in particular those with shorter rods,
have even smaller order parameters.

\begin{figure}[ht]
  \includegraphics[width=0.5\textwidth]{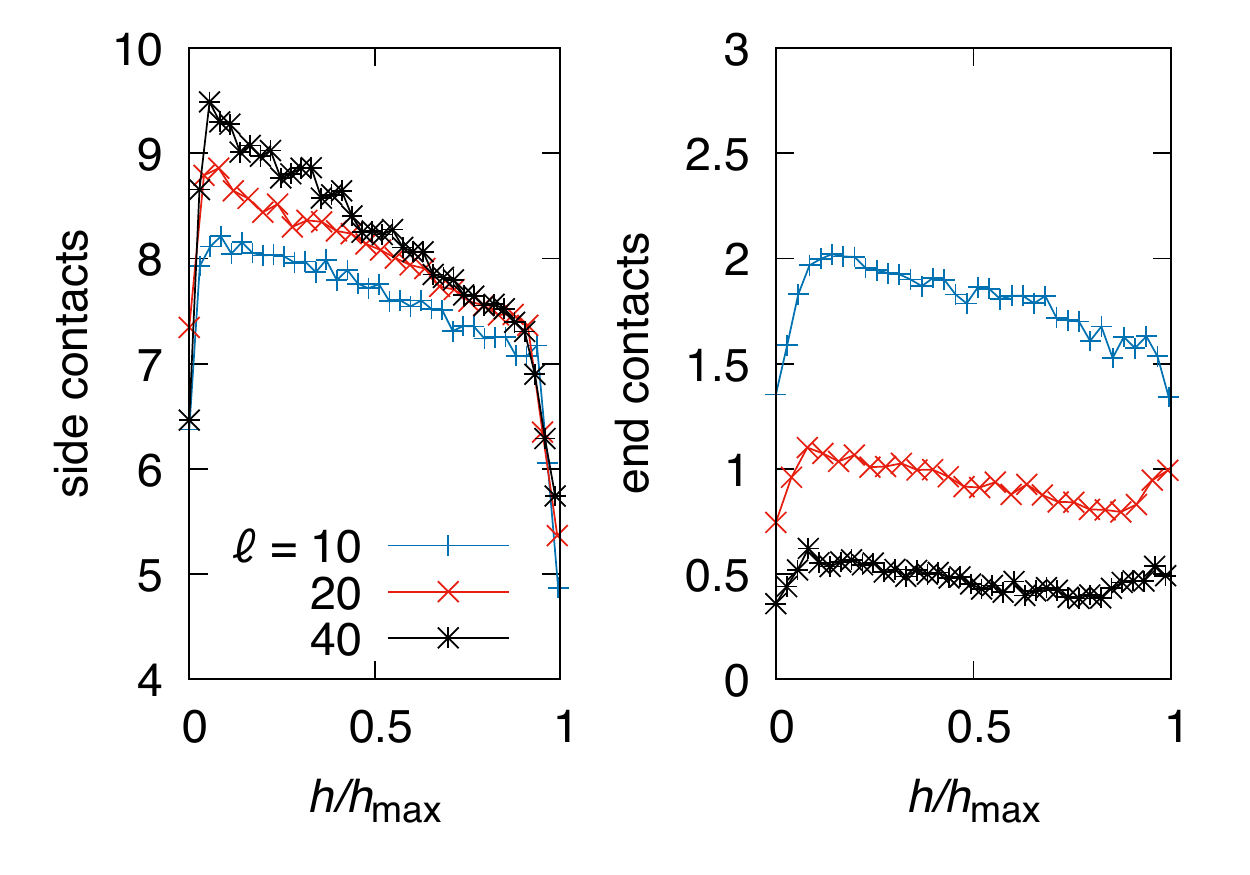}
  \caption{Average number of contacts per particle vs. normalized
    height $h/h_{\rm max}$ for columns with different $\ell$ before
    removal of the wall; averaged over columns with different $\phi_0$;
    $\mu=0.1$.}
  \label{fig:zend}
\end{figure}

We have also measured the number and nature of inter-particle
contacts. A contact may occur either at the side or at the end of a
spherocylinder. Fig.~\ref{fig:zend} displays these two different types
separately and resolves the height-dependence within the column. The
maximum height $h_{\rm max}$ of the column is determined by the
center-of-mass of the topmost spherocylinder.
The number of side contacts increases with depth, $h\to0$. The effect
is stronger the longer the particles are. Towards the upper end all
three columns have about the same number of side contacts per
particle, approx. $7-7.5$. In Ref.~\cite{PhysRevE.103.052903} a
minimal value of $6-6.5$ has been found for an isotropically jammed
(i.e. without gravity) frictional system. In isotropic frictionless
systems the number of contacts is $8+2f$, where $f$ is the fraction of
rods that have end contacts at both
ends~\cite{PhysRevE.102.022903}\footnote{The reason for the value
  $8+2f=2(4+f)$ is constraint counting: every rod has 4 degrees of
  freedom (excluding rotations around and translations along the long
  axis). The longitudinal translation is only counted for the $f$ rods
  that are longitudinally constrained via end contacts on both
  ends.}.

The overall number of end contacts per
rod is displayed in the right panel. As to be expected it is much
smaller than the number of side contacts. What is more, and different
to the number of side contacts, the number of end contacts strongly
decreases with particle length, approximately as $\propto1/\ell$.

\begin{figure}[t]
  \includegraphics[width=0.45\textwidth]{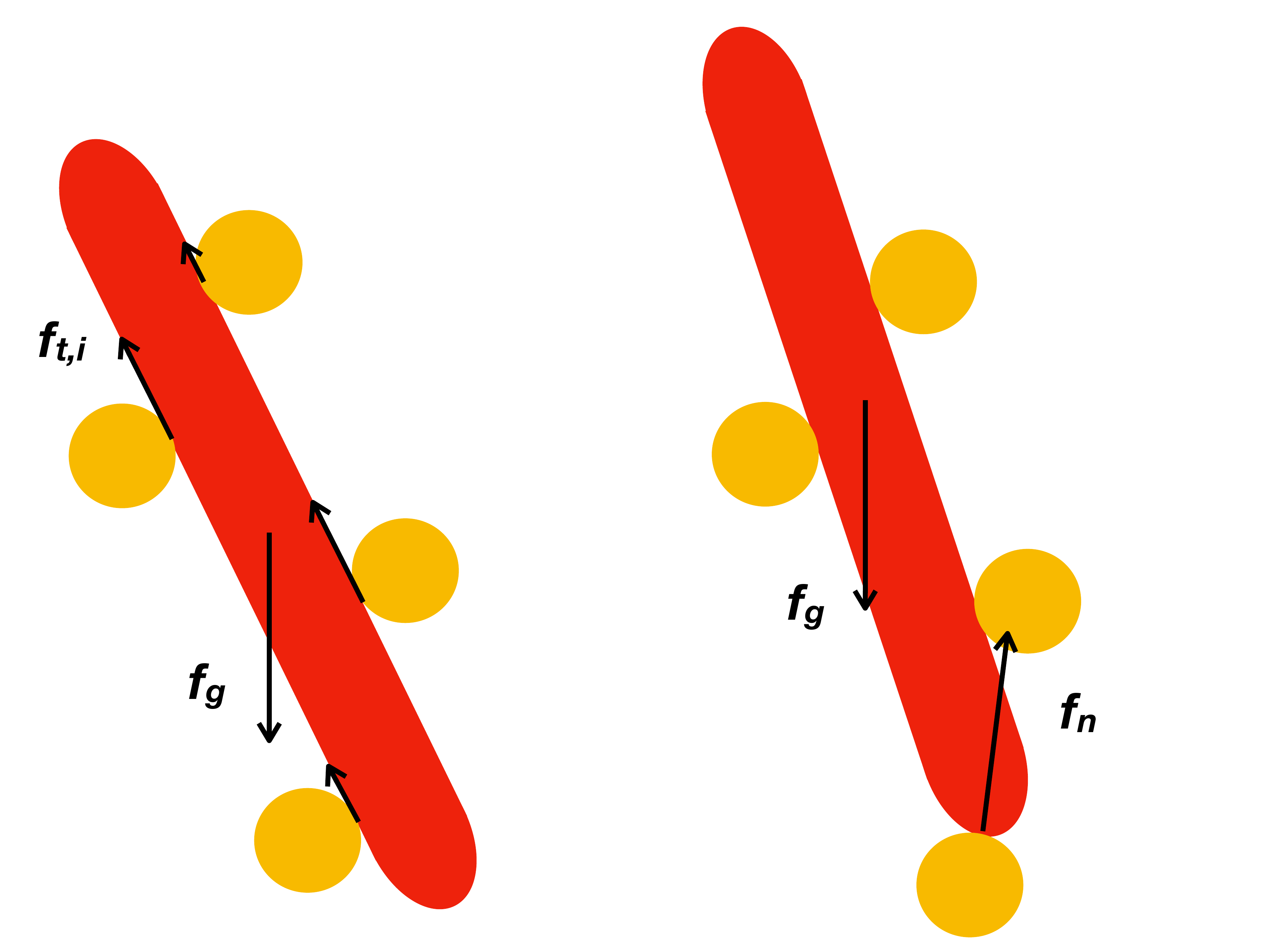}
  \caption{Illustration of the two different ways to stabilize a rod
    that slides downwards within the column. Gravitational force may
    either be competed by the frictional forces $f_{t,i}$ from the
    side-contacts $i$, or by the steric force (normal force $f_n$)
    from one (or several) contacts at the rod tip. Orange disks
    represent surrounding rods that serve as obstacles; they can be
    thought of as rods pointing out-of-plane. }
  \label{fig:sketch_obstacle}
\end{figure}

End contacts may play an important role in putting a stop to the
process of compaction of the column in the gravitational field. A
higher number of end contacts then implies a smaller tendency to
compaction, in line with the observations reported in
Table~\ref{tbl:rel_compaction}. During gravitational compaction rods
slide downwards within the packing of other rods. The surrounding is
dense, rotations are frozen out and motion is primarily along the long
axis.  On its way the rod may collide head-on with a rod that is
already part of a stable structure. If the obstacle cannot be bypassed
it will come to rest via steric hindrance (see
Fig.~\ref{fig:sketch_obstacle} right). In consequence a stable end
contact is formed. The higher the occupied volume the more likely it
its for the rod to undergo such a collision, $z_e \propto \phi\sim
\ell^{-1}$, as approximately observed in Fig.~\ref{fig:zend}.

Particles in packings with fewer end contacts also come to rest
eventually (take for example the fraction of particles with an
end-contact only at the upper end). Fig.~\ref{fig:sketch_obstacle}
illustrates a second possibility how rods can be stabilized against
gravitational forces within the packing. As compaction proceeds, rods
squeeze through ever narrower pores and start to feel more and more
contacts at their sides. These contacts produce upward forces from
friction. A rod, and thus compaction, can finally come to rest when
these frictional contact forces from side contacts are able to surpass
the gravitational force pushing downwards.

\begin{figure}[ht]
  \includegraphics[width=0.5\textwidth]{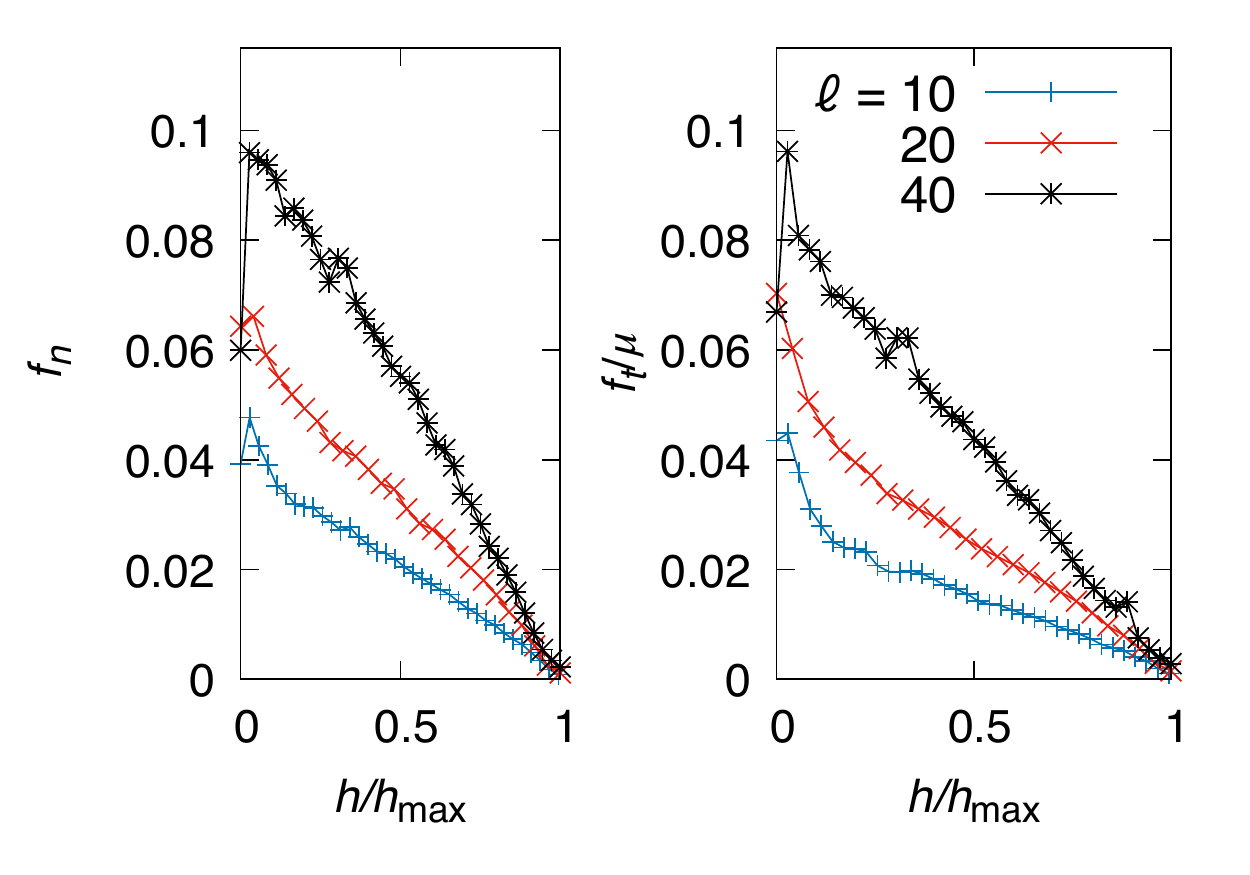}
  \caption{Average contact forces $f_{n/t}$ vs. normalized height
    $h/h_{\rm max}$ for columnss with different $\ell$ just before
    removal of the wall; $\mu=0.1$.}
  \label{fig:fn.z}
\end{figure}

It is illuminating to also plot these contact forces, frictional and
normal forces for the different columns (Fig.~\ref{fig:fn.z}). As
expected the forces increase linearly with depth, $h\to0$. The scale
is the weight from the material above, $\sim mg(1-h/h_{\rm max})$.

The overall magnitude of the average normal contact force $f_n$ and
the frictional force $f_t/\mu$ are of the same order of magnitude,
with the frictional forces somewhat smaller. The Coulomb threshold is
therefore generally not reached and contacts are not sliding in this
compacted state.

What is most prominent, however, is the trend with $\ell$, similar to
what is observed for the side contacts.  Forces are larger for longer
rods. For the three values of $\ell=10,20,40$ we approximately find
$f_{n,t}\propto \ell$. This is to be compared with the behavior of the
pressure, which does not depend on $\ell$. Being only governed by the
mass distribution, the vertical pressure very closely follows the
expected law $p_v = \rho_mg(h_{\rm max}-h)$.

In conclusion, packings with long rods have a large relative
compaction. The packing stabilizes and comes to rest when particles
feel enough confining force from contacts at their sides. The
associated frictional forces point upwards and can exceed the downward
gravitational force. Packings with shorter rods have a smaller
relative compaction, because particles falling down within the column
have a higher propability (larger density) to encounter obstacles at
their tips. This may stabilize the rod and stop further settling.

\subsection{Dynamics after removal of walls}

After the column has come to rest within the container the walls in
y-direction are removed and the column starts to yield. The associated
process for spherical grains has been studied in dozens of
publications. The mechanism for failure is quite simple: the weight
from particles above leads to horizontal outward forces in the lower
parts of the column. After removal of the container walls these forces
can no longer be compensated such that the column starts to
yield. This happens by squeezing apart the lower layers, while the
upper layers start to fall due to the gravitational acceleration, this
way keeping up the compression on the lower layers. The friction
coefficient does not have much influence on this process. Increasing
friction might have some effect on the final run-out length via an
increased amount of energy dissipation. But in keeping the column
together and stable, frictional forces are not at all helpful.

\begin{figure}[t]
  \includegraphics[width=0.4\textwidth]{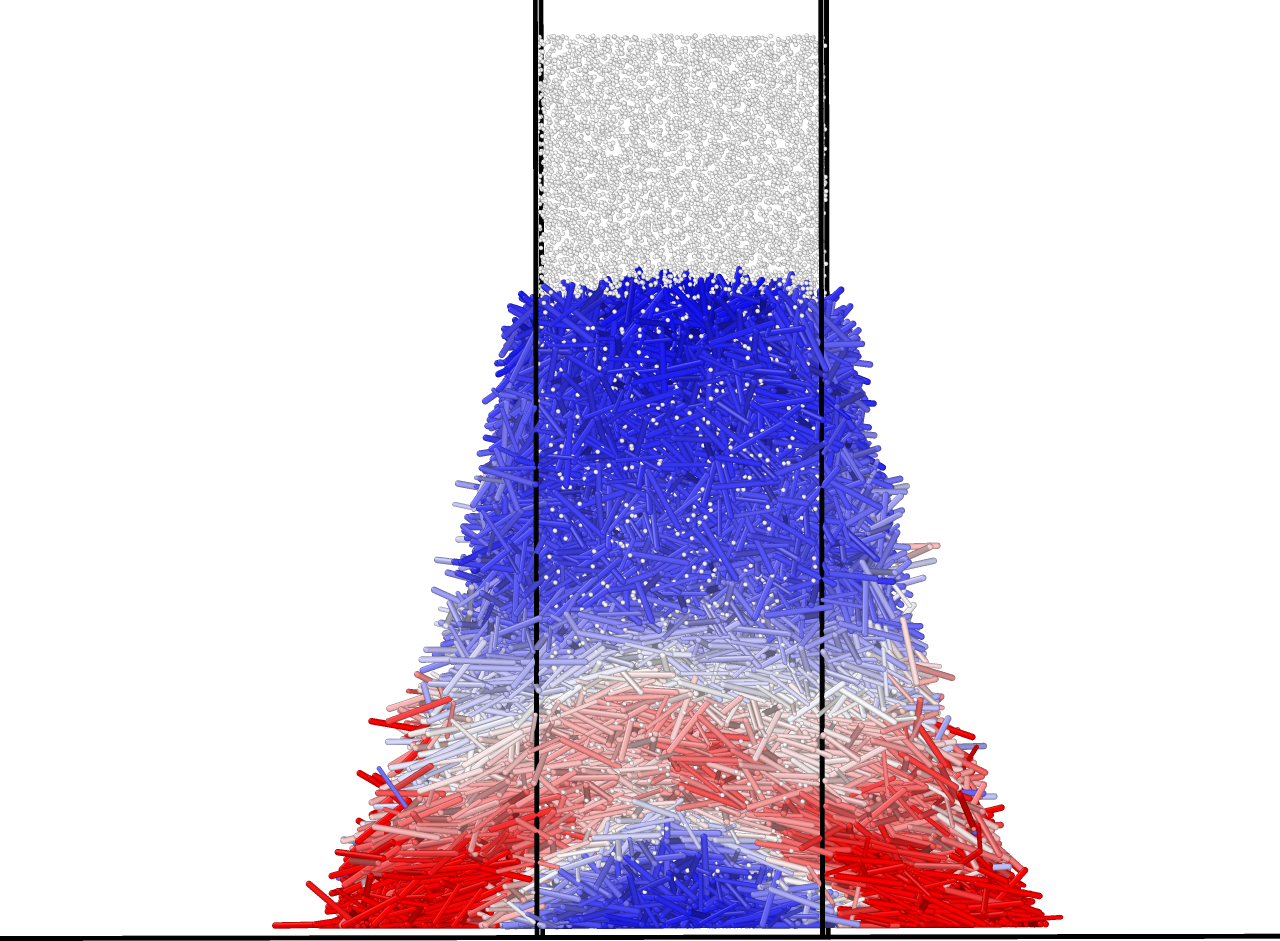}
  \caption{Sideview (yz-plane) of a column with $\ell=10$ and
    $\mu=0.1$ at time $t=300$, which corresponds roughly to the
    gravitational time-scale $t(l)=\sqrt{2l/g}$ on the scale of one
    third of the column height, $l\approx L_z/3$. Particles are
    colored according to local shear strain (blue-small,
    red-large). Gray points represent center of masses of rods just
    before collapse.}
  \label{fig:collapse10}
\end{figure}

Fig.\ref{fig:collapse10} illustrates that columns of shorter rods at
small enough friction coefficient fail in a manner similar to columns
of spherical particles. The color-code highlights the zones of maximal
strain to be at the bottom. Here, the column presses outwards, while
the upper part is only weakly deformed during its phase of nearly-free
fall. There is also a stagnant zone in the center right above the
surface. Particles from there don't move much.

The behavior described here, represents the column collapse far away
from the transition as described in Figs.~\ref{fig:frac_mu_ell} and
\ref{fig:phase_mu_ell}. By increasing friction the transition line is
approached and the collapse changes qualitatively. Eventually, with
large enough friction coefficient, the column does not collapse
anymore, but stays solid-like. Apparently, and different to the case
of spheres, friction (as encoded in $\mu$) here is a relevant
variable.  Frictional forces \emph{are} able to provide the necessary
``cohesion'' to oppose the destabilizing forces of gravity.

\begin{figure*}[t]
  \includegraphics[width=0.3\textwidth]{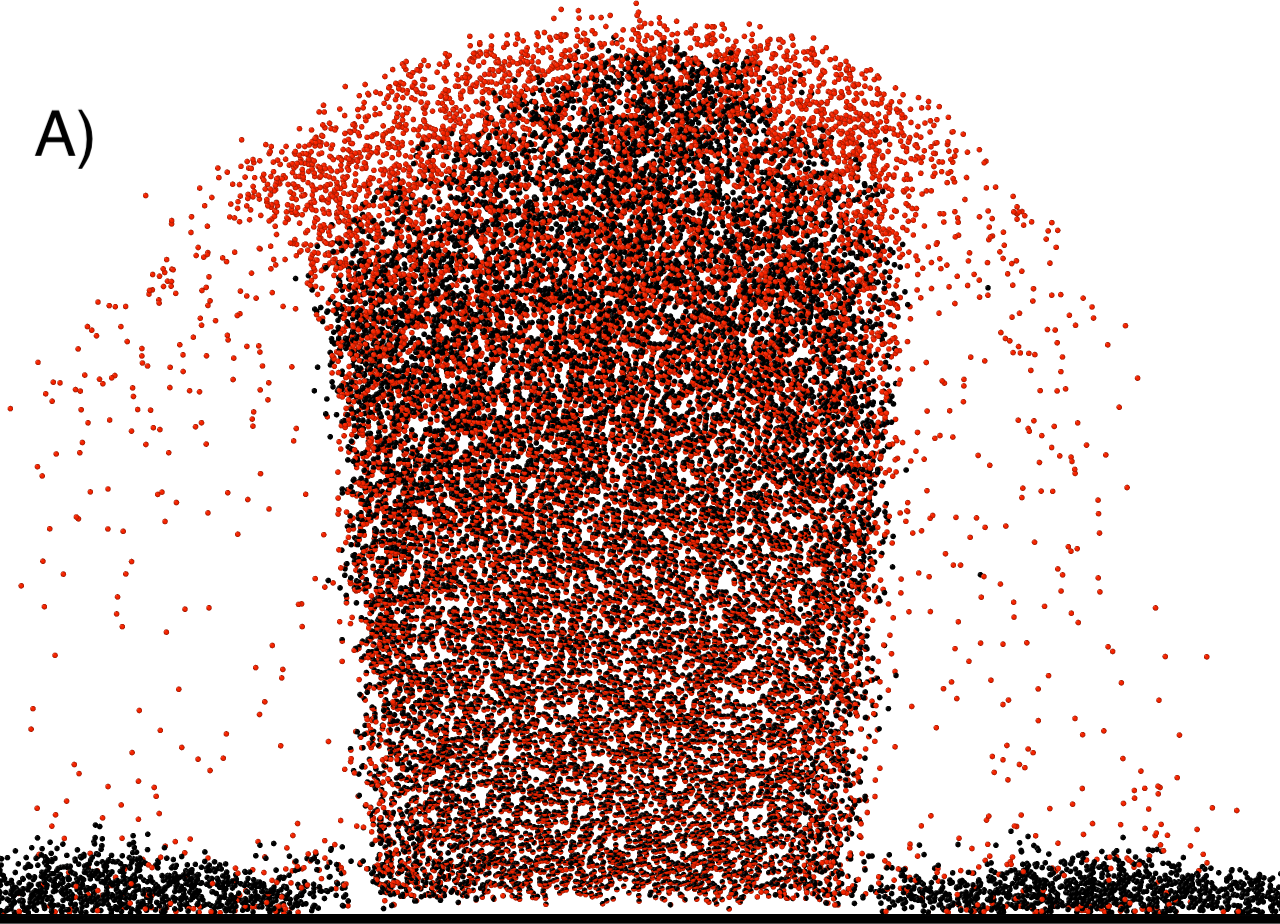}
  \includegraphics[width=0.3\textwidth]{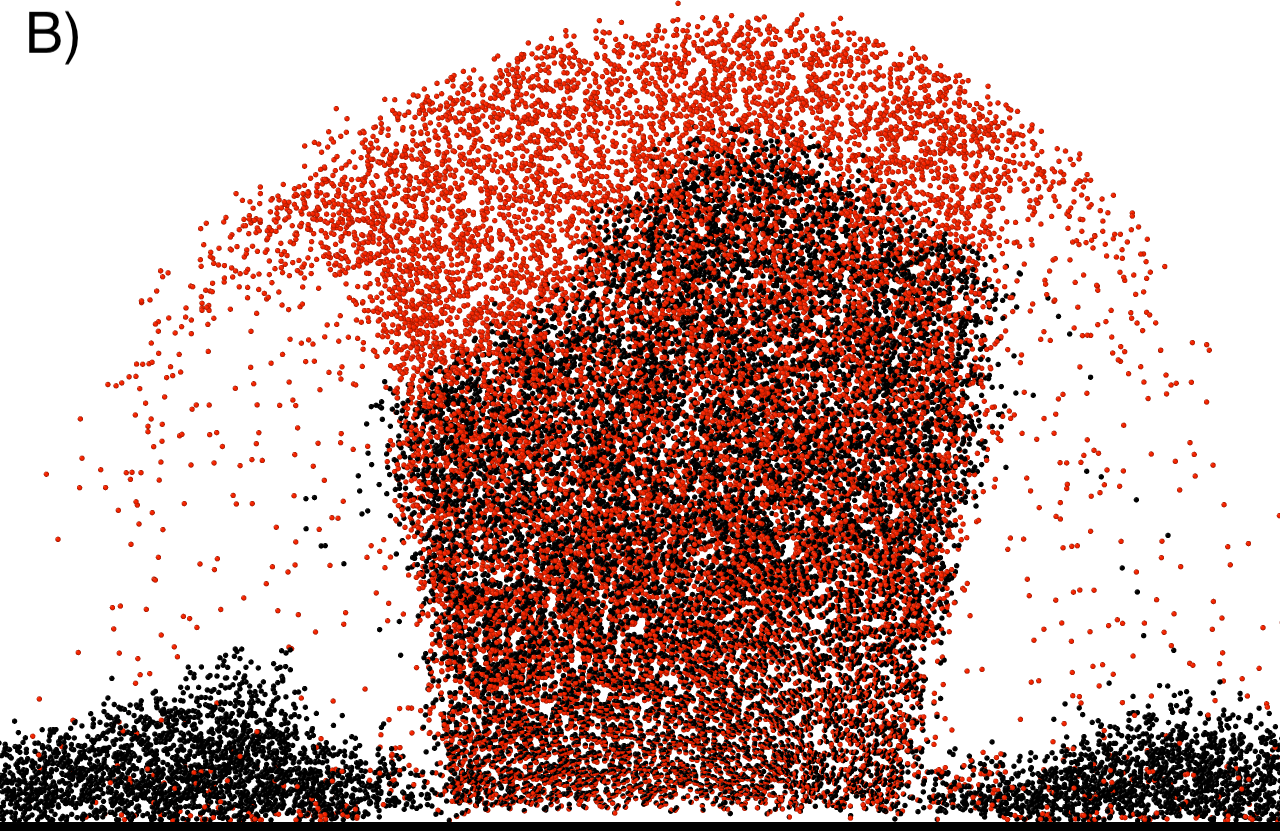}
  \includegraphics[width=0.3\textwidth]{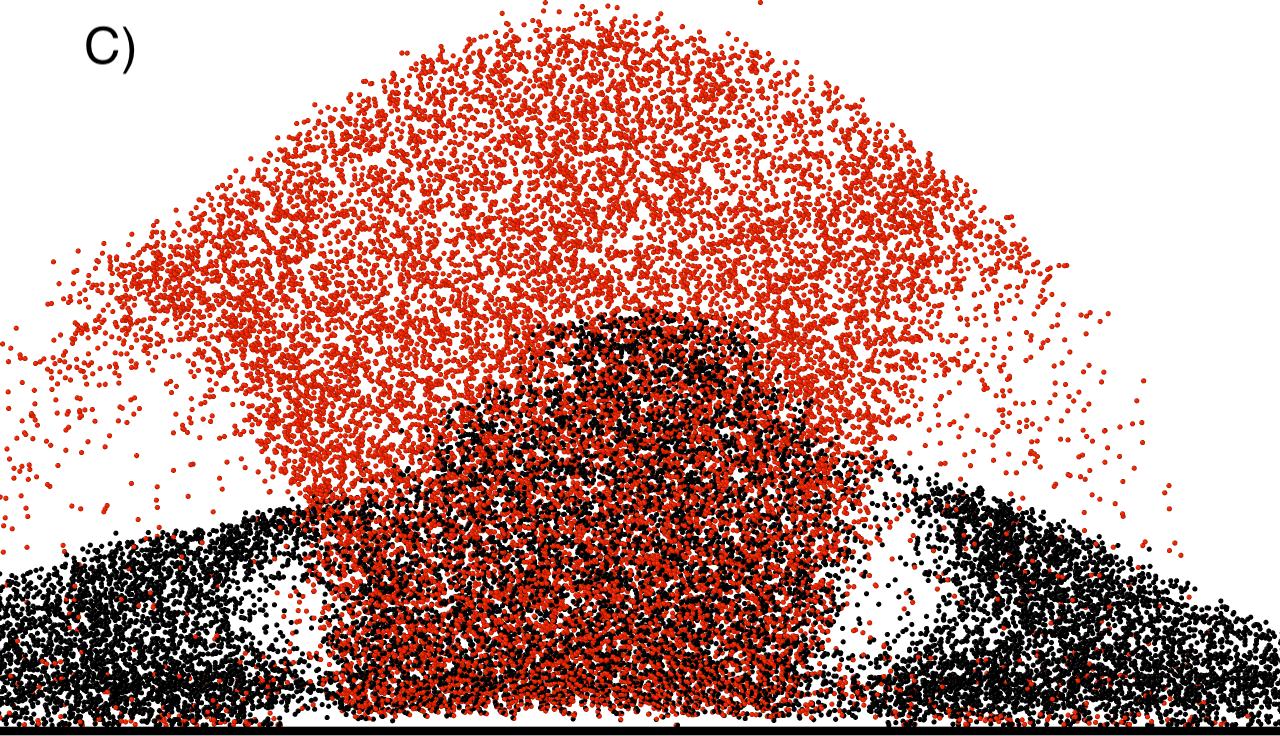}
  \caption{Sideview (yz-plane) of a column with $\ell=40$ and different
    $\mu$ close to $\mu_c(\ell)$; comparison of intermediate (red,
    time $t=1000$) and late (black, $t=20000$) stage of collapse;
    displayed are the center of masses of the particles: (A)
    $\mu=0.08$; (B) $\mu=0.07$ (C) $\mu=0.06$.}
  \label{fig:snapshot_com}
\end{figure*}

Details about the collapse behavior close to the transition are given
in Fig.~\ref{fig:snapshot_com}, where columns with different $\mu$
(decreasing from left to right) close to the transition
$\mu_c(\ell=40)$ are displayed at intermediate and at late times. To
ease visualization only the center-of-masses of the particles are
displayed, and not the entire spherocylindrical form as in
Fig.~\ref{fig:snapshots_stable}.

Apparently, the columns are quite stable at the bottom, keeping
together there. The strong overlap between red and black points (in
panels A) and B)) highlights that particles in the lower part of the
column hardly move between the two snapshots. The column is solid
there.

Particles fall down mainly from the top. A stable free surface is
developed, interestingly with an overhang. For the lowest $\mu=0.06$
(panel C)) the surface is stable only at intermediate times, while an
avalanche of rods slide down from the top, eroding the column ever
more. In the final state (black), the core part of the remaining
column is stabilized by the deposit of particles that have been eroded from
the top of the column. Interesting is also the depletion zone between
core and deposit, which arises because of an umbrella-effect of the
overhanging surface. This depletion zone is only hardly visible when
displaying the entire spherocylinders instead of only the center of
masses.

The time-scale of the collapse here is also much longer than in
Fig.~\ref{fig:collapse10}. The column depicted there, would already
have come to rest in its final shape at the intermediate time $t=1000$
(red).

Increasing friction thus leads to a growing solid-like region from
bottom to top. For small friction it is only the small zone visible in
Fig.~\ref{fig:collapse10} that does not flow. Increasing friction,
larger zones remain solid and particle flow happens only further up
towards the top of the column. When the friction is large enough to
solidify the entire column, only individual particles from the top or
the sides slide off and fall down one after the other.

Why is it possible that packings of rods can utilize frictional forces
to oppose the collapse of the column, while packings of spherical
particles cannot? A contact between two rods is established, where the
shortest distance between the backbones of two particles is
achieved. The contact vector $\mathbf{r}_{ij}$ is, in general, not
correlated with the vector between center-of masses $\mathbf{R}_{ij}$,
\begin{equation}\label{eq:Rij}
  \mathbf{r}_{ij} = \mathbf{R}_{ij}+\mathbf{s}_i - \mathbf{s}_j\,,
\end{equation}
where the ``arclength'' vectors $\mathbf{s}_{i,j}$ point from the
center of mass to the position along the backbone, where the contact
is established. These are zero in spheres, such that both vectors
$\mathbf{r}_{ij}$ and $\mathbf{R}_{ij}$ are identical.

Spheres that undergo dilational strain quickly loose contact, rather
than building-up proper frictional forces that resist the
strain. Quite different with rod-like particles, where a changing
$\mathbf{R}_{ij}$ can be accomodated by changing $\mathbf{s}_{i,j}$,
that is the contact may persist by moving tangentially along the
backbones of the particles. This motion builds up the proper
frictional forces to oppose lateral spreading.

\begin{figure}[ht]
  \includegraphics[width=0.5\textwidth]{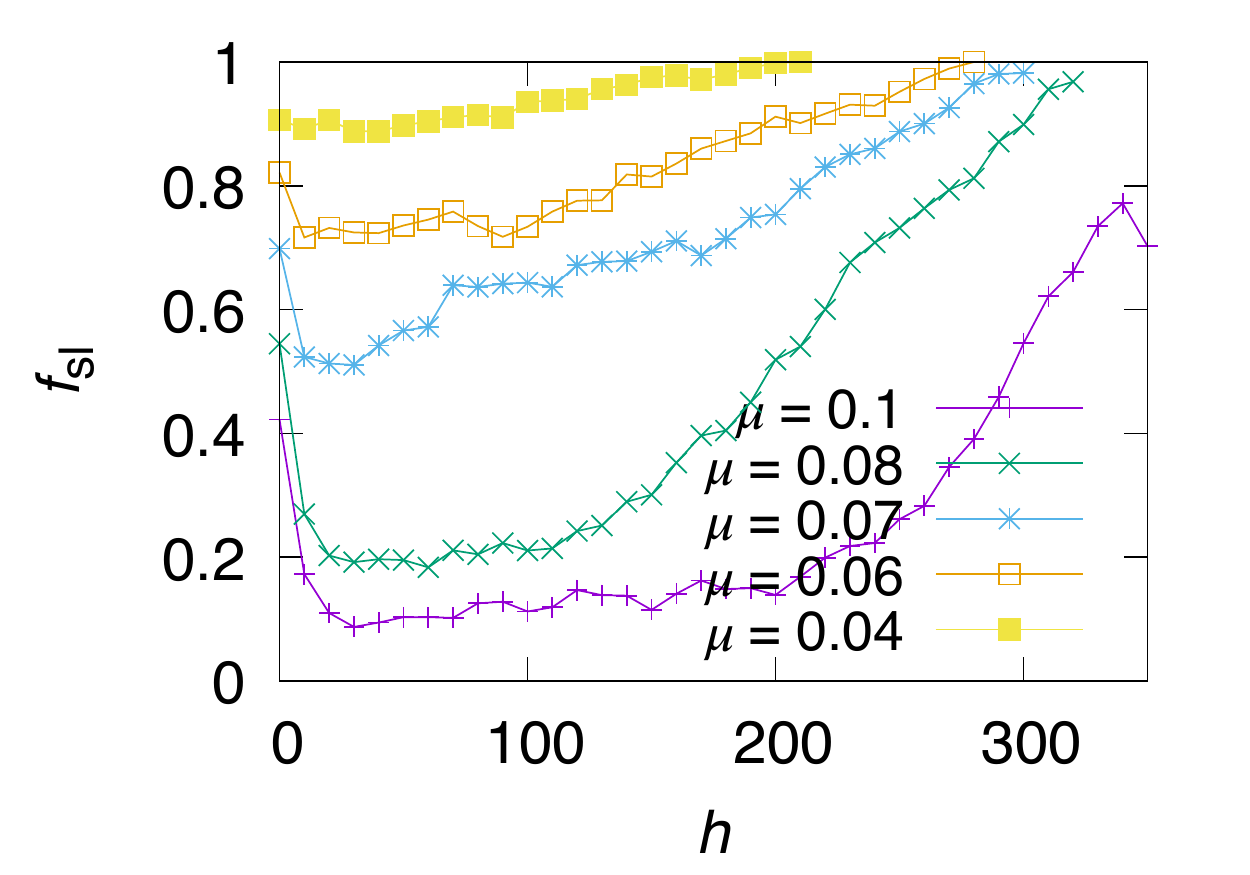}
  \caption{Fraction of sliding contacts $f_{\rm sl}$ vs. height $z$
    for different $\mu$. Same columns as in Fig.~\ref{fig:snapshot_com}
    (and in addition those from Figs.~\ref{fig:snapshots_stable} and
    \ref{fig:snapshots_unstable}); taken at the intermediate time
    $t=1000$.}
  \label{fig:sliding_contacts}
\end{figure}

The role of $\mu$ is to represent the maximal frictional force of a
contact, via the Coulomb inequality. Once the maximum is reached, the
contact is sliding and cannot contribute any further to the
stabilization of the packing. A large number of sliding contacts is
therefore a clear sign of an unstable
column. Fig.~\ref{fig:sliding_contacts} displays the fraction of
sliding contacts $f_{\rm sl}$ resolved according to height $h$. The
stable columns at higher $\mu$ have a very small fraction of sliding
contacts. The fraction increases with decreasing $\mu$. In the upper
part of all columns many contacts are sliding, in agreement with the
erosion processes occuring there.

In conclusion, column stability is mediated by the number of non-sliding
contacts. In contrast to sphere-contacts, SC-contacts can be
long-lived by moving tangentially along SC sides to mobilize
additional frictional forces necessary to oppose gravity-induced
destabilizing forces. Higher values of $\mu$ imply higher frictional
forces and thus better stability for the column.

\subsection{Conclusion}

This work discusses the stability of free-standing columns of granular
materials in gravity. Particles have the form of long thin rods. After
preparing the column within a container, the walls are removed to
generate free vertical surfaces. Depending on the conditions we
observe full collapse of the column or indeed the establishment of a
quasi-solid free standing column with stable vertical, or even
overhanging, surfaces. Such an effect has previously been observed in
experiments, Ref.\cite{desmond06:_jammin,PhysRevE.82.011308}, however
no simulations exist up to date to study this effect.

Friction angles of the ``usual'' granular particles are far below the
vertical $90^\circ$. Similarly, we observe that columns usually
collapse into shallow heaps if the rods are short enough to qualify as
``nearly-spherical'', or if the friction coefficient is small enough.

In a granular column the vertical, gravity-induced pressure induces
horizontal pressure components which are balanced by the container
walls. After removal of the walls these forces need to be balanced by
internal ``cohesive'' forces in order to prevent full collapse of the
structure. In our packings of rods frictional forces are built up that
indeed are able to oppose the destabilizing forces of gravity. This is
possible because inter-rod contacts, as opposed to inter-sphere
contacts, may persist for larger (even dilational) strains, just by
moving along the sides of the rods.

The maximal friction force on a contact is set by the Coulomb
condition $f_t=\mu f_n$. A higher friction coefficient $\mu$ then
implies more stability for the column, because larger friction forces
are possible. We also find that longer particles ($\ell$) give more
stable columns, even though friction coefficients may be small.  This is
because longer rods imply higher normal forces $f_n$, which also
increases the maximal friction force. Inter-particle forces are
related to pressure via $p\sim \rho z\langle f_nR\rangle$. With
densities $\rho\sim\ell^{-2}$ because of excluded volume, and center
of mass distances $R\sim\ell$, forces can be written as $f\sim
pz\ell$, and thus increase with $\ell$.

We also analyze the packed state within the container before removing
the sidewall. Most striking is the lack of contacts at rod ends in
columns with long rods; only every second rod has on average an end
contact when $\ell=40$, while rods with $\ell=20$ have one contact per
rod on average. On the other hand, the number of side contacts is
largely independent of length.

The two types of contacts highlight two possible mechanisms to achieve
stability of a rod within a nearly jammed packing of other rods. In
this dense surrounding rotations are frozen out and motion is
primarily along the long axis. During gravitational compaction the rod
may collide head-on with a rod that is already part of a stable
structure. If the obstacle cannot be bypassed it will come to rest via
steric hindrance - and form a stable end contact. Such a mechanism is
more likely in packings of short rods where the overall density is
high and where it is more likely to encounter a second rod to
establish a new end-constraint. In packings of long rods overall
density is small (approximately, $\rho\propto \ell^{-2}$) and end
contacts are less likely to occur. Rather, rod stabilization comes
from the confining forces from contacts at its side. It is the
associated frictional forces, acting tangentially along the rod axis
that may stop further sliding downwards.

\begin{acknowledgments}

  I acknowledge financial support by the German Science Foundation
  (Deutsche Forschungsgemeinschaft) via the Heisenberg program
  (HE-6322/2).
  
\end{acknowledgments}

%\bibliography{../../notes/references,/Users/claus/ownCloud/bib_files/references.all,../cite-paper}

\begin{thebibliography}{38}%
\makeatletter
\providecommand \@ifxundefined [1]{%
 \@ifx{#1\undefined}
}%
\providecommand \@ifnum [1]{%
 \ifnum #1\expandafter \@firstoftwo
 \else \expandafter \@secondoftwo
 \fi
}%
\providecommand \@ifx [1]{%
 \ifx #1\expandafter \@firstoftwo
 \else \expandafter \@secondoftwo
 \fi
}%
\providecommand \natexlab [1]{#1}%
\providecommand \enquote  [1]{``#1''}%
\providecommand \bibnamefont  [1]{#1}%
\providecommand \bibfnamefont [1]{#1}%
\providecommand \citenamefont [1]{#1}%
\providecommand \href@noop [0]{\@secondoftwo}%
\providecommand \href [0]{\begingroup \@sanitize@url \@href}%
\providecommand \@href[1]{\@@startlink{#1}\@@href}%
\providecommand \@@href[1]{\endgroup#1\@@endlink}%
\providecommand \@sanitize@url [0]{\catcode `\\12\catcode `\$12\catcode
  `\&12\catcode `\#12\catcode `\^12\catcode `\_12\catcode `\%12\relax}%
\providecommand \@@startlink[1]{}%
\providecommand \@@endlink[0]{}%
\providecommand \url  [0]{\begingroup\@sanitize@url \@url }%
\providecommand \@url [1]{\endgroup\@href {#1}{\urlprefix }}%
\providecommand \urlprefix  [0]{URL }%
\providecommand \Eprint [0]{\href }%
\providecommand \doibase [0]{http://dx.doi.org/}%
\providecommand \selectlanguage [0]{\@gobble}%
\providecommand \bibinfo  [0]{\@secondoftwo}%
\providecommand \bibfield  [0]{\@secondoftwo}%
\providecommand \translation [1]{[#1]}%
\providecommand \BibitemOpen [0]{}%
\providecommand \bibitemStop [0]{}%
\providecommand \bibitemNoStop [0]{.\EOS\space}%
\providecommand \EOS [0]{\spacefactor3000\relax}%
\providecommand \BibitemShut  [1]{\csname bibitem#1\endcsname}%
\let\auto@bib@innerbib\@empty
%</preamble>
\bibitem [{\citenamefont {Ashour}\ \emph {et~al.}(2017)\citenamefont {Ashour},
  \citenamefont {Wegner}, \citenamefont {Trittel}, \citenamefont
  {B{\"o}rzs{\"o}nyi},\ and\ \citenamefont {Stannarius}}]{ashour17:_outfl}%
  \BibitemOpen
  \bibfield  {author} {\bibinfo {author} {\bibfnamefont {A.}~\bibnamefont
  {Ashour}}, \bibinfo {author} {\bibfnamefont {S.}~\bibnamefont {Wegner}},
  \bibinfo {author} {\bibfnamefont {T.}~\bibnamefont {Trittel}}, \bibinfo
  {author} {\bibfnamefont {T.}~\bibnamefont {B{\"o}rzs{\"o}nyi}}, \ and\
  \bibinfo {author} {\bibfnamefont {R.}~\bibnamefont {Stannarius}},\
  }\href@noop {} {\bibfield  {journal} {\bibinfo  {journal} {Soft Matter}\
  }\textbf {\bibinfo {volume} {13}},\ \bibinfo {pages} {402} (\bibinfo {year}
  {2017})}\BibitemShut {NoStop}%
\bibitem [{\citenamefont {Kroy}\ \emph {et~al.}(2002)\citenamefont {Kroy},
  \citenamefont {Sauermann},\ and\ \citenamefont
  {Herrmann}}]{kroy02:_minim_model_sand_dunes}%
  \BibitemOpen
  \bibfield  {author} {\bibinfo {author} {\bibfnamefont {K.}~\bibnamefont
  {Kroy}}, \bibinfo {author} {\bibfnamefont {G.}~\bibnamefont {Sauermann}}, \
  and\ \bibinfo {author} {\bibfnamefont {H.~J.}\ \bibnamefont {Herrmann}},\
  }\href@noop {} {\bibfield  {journal} {\bibinfo  {journal} {Phys. Rev. Lett.}\
  }\textbf {\bibinfo {volume} {88}},\ \bibinfo {pages} {054301} (\bibinfo
  {year} {2002})}\BibitemShut {NoStop}%
\bibitem [{\citenamefont {Al-Hashemi}\ and\ \citenamefont
  {Al-Amoudi}(2018)}]{al-hashemi18}%
  \BibitemOpen
  \bibfield  {author} {\bibinfo {author} {\bibfnamefont {H.~M.~B.}\
  \bibnamefont {Al-Hashemi}}\ and\ \bibinfo {author} {\bibfnamefont {O.~S.~B.}\
  \bibnamefont {Al-Amoudi}},\ }\href@noop {} {\bibfield  {journal} {\bibinfo
  {journal} {Powder Technology}\ }\textbf {\bibinfo {volume} {330}},\ \bibinfo
  {pages} {397} (\bibinfo {year} {2018})}\BibitemShut {NoStop}%
\bibitem [{\citenamefont {Lacaze}\ \emph {et~al.}(2008)\citenamefont {Lacaze},
  \citenamefont {Phillipse},\ and\ \citenamefont
  {Kerswell}}]{lacaze08:_planar}%
  \BibitemOpen
  \bibfield  {author} {\bibinfo {author} {\bibfnamefont {L.}~\bibnamefont
  {Lacaze}}, \bibinfo {author} {\bibfnamefont {J.~C.}\ \bibnamefont
  {Phillipse}}, \ and\ \bibinfo {author} {\bibfnamefont {R.~R.}\ \bibnamefont
  {Kerswell}},\ }\href@noop {} {\bibfield  {journal} {\bibinfo  {journal}
  {Phys. Fluids}\ }\textbf {\bibinfo {volume} {20}},\ \bibinfo {pages} {063302}
  (\bibinfo {year} {2008})}\BibitemShut {NoStop}%
\bibitem [{\citenamefont {Lagr{\'e}e}\ \emph {et~al.}(2011)\citenamefont
  {Lagr{\'e}e}, \citenamefont {Staron},\ and\ \citenamefont
  {Popinet}}]{lagree11}%
  \BibitemOpen
  \bibfield  {author} {\bibinfo {author} {\bibfnamefont {P.-Y.}\ \bibnamefont
  {Lagr{\'e}e}}, \bibinfo {author} {\bibfnamefont {L.}~\bibnamefont {Staron}},
  \ and\ \bibinfo {author} {\bibfnamefont {S.}~\bibnamefont {Popinet}},\
  }\href@noop {} {\bibfield  {journal} {\bibinfo  {journal} {J. Fluid Mech.}\
  }\textbf {\bibinfo {volume} {686}},\ \bibinfo {pages} {378} (\bibinfo {year}
  {2011})}\BibitemShut {NoStop}%
\bibitem [{\citenamefont {Holsapple}(2013)}]{holsapple13:_model}%
  \BibitemOpen
  \bibfield  {author} {\bibinfo {author} {\bibfnamefont {K.~A.}\ \bibnamefont
  {Holsapple}},\ }\href@noop {} {\bibfield  {journal} {\bibinfo  {journal}
  {Planetary and Space Science}\ }\textbf {\bibinfo {volume} {82-83}},\
  \bibinfo {pages} {11} (\bibinfo {year} {2013})}\BibitemShut {NoStop}%
\bibitem [{\citenamefont {Xu}\ \emph {et~al.}(2016)\citenamefont {Xu},
  \citenamefont {Sun}, \citenamefont {Jin},\ and\ \citenamefont
  {Chen}}]{xu16:_measur}%
  \BibitemOpen
  \bibfield  {author} {\bibinfo {author} {\bibfnamefont {X.}~\bibnamefont
  {Xu}}, \bibinfo {author} {\bibfnamefont {Q.}~\bibnamefont {Sun}}, \bibinfo
  {author} {\bibfnamefont {F.}~\bibnamefont {Jin}}, \ and\ \bibinfo {author}
  {\bibfnamefont {Y.}~\bibnamefont {Chen}},\ }\href@noop {} {\bibfield
  {journal} {\bibinfo  {journal} {Powder Technology}\ }\textbf {\bibinfo
  {volume} {303}},\ \bibinfo {pages} {147} (\bibinfo {year}
  {2016})}\BibitemShut {NoStop}%
\bibitem [{\citenamefont {Lube}\ \emph {et~al.}(2007)\citenamefont {Lube},
  \citenamefont {Huppert}, \citenamefont {Sparks},\ and\ \citenamefont
  {Freundt}}]{lube07:_static}%
  \BibitemOpen
  \bibfield  {author} {\bibinfo {author} {\bibfnamefont {G.}~\bibnamefont
  {Lube}}, \bibinfo {author} {\bibfnamefont {H.~E.}\ \bibnamefont {Huppert}},
  \bibinfo {author} {\bibfnamefont {R.~S.~J.}\ \bibnamefont {Sparks}}, \ and\
  \bibinfo {author} {\bibfnamefont {A.}~\bibnamefont {Freundt}},\ }\href@noop
  {} {\bibfield  {journal} {\bibinfo  {journal} {Phys. Fluids}\ }\textbf
  {\bibinfo {volume} {19}},\ \bibinfo {pages} {043301} (\bibinfo {year}
  {2007})}\BibitemShut {NoStop}%
\bibitem [{\citenamefont {Lajeunesse}\ \emph {et~al.}(2004)\citenamefont
  {Lajeunesse}, \citenamefont {Mangeney-Castelnau},\ and\ \citenamefont
  {Vilotte}}]{lajeunesse04:_spread}%
  \BibitemOpen
  \bibfield  {author} {\bibinfo {author} {\bibfnamefont {E.}~\bibnamefont
  {Lajeunesse}}, \bibinfo {author} {\bibfnamefont {A.}~\bibnamefont
  {Mangeney-Castelnau}}, \ and\ \bibinfo {author} {\bibfnamefont {J.~P.}\
  \bibnamefont {Vilotte}},\ }\href@noop {} {\bibfield  {journal} {\bibinfo
  {journal} {Phys. Fluids}\ }\textbf {\bibinfo {volume} {16}},\ \bibinfo
  {pages} {2371} (\bibinfo {year} {2004})}\BibitemShut {NoStop}%
\bibitem [{\citenamefont {Torquato}\ and\ \citenamefont
  {Stillinger}(2010)}]{RevModPhys.82.2633}%
  \BibitemOpen
  \bibfield  {author} {\bibinfo {author} {\bibfnamefont {S.}~\bibnamefont
  {Torquato}}\ and\ \bibinfo {author} {\bibfnamefont {F.~H.}\ \bibnamefont
  {Stillinger}},\ }\href {\doibase 10.1103/RevModPhys.82.2633} {\bibfield
  {journal} {\bibinfo  {journal} {Rev. Mod. Phys.}\ }\textbf {\bibinfo {volume}
  {82}},\ \bibinfo {pages} {2633} (\bibinfo {year} {2010})}\BibitemShut
  {NoStop}%
\bibitem [{\citenamefont {VanderWerf}\ \emph {et~al.}(2018)\citenamefont
  {VanderWerf}, \citenamefont {Jin}, \citenamefont {Shattuck},\ and\
  \citenamefont {O'Hern}}]{PhysRevE.97.012909}%
  \BibitemOpen
  \bibfield  {author} {\bibinfo {author} {\bibfnamefont {K.}~\bibnamefont
  {VanderWerf}}, \bibinfo {author} {\bibfnamefont {W.}~\bibnamefont {Jin}},
  \bibinfo {author} {\bibfnamefont {M.~D.}\ \bibnamefont {Shattuck}}, \ and\
  \bibinfo {author} {\bibfnamefont {C.~S.}\ \bibnamefont {O'Hern}},\ }\href
  {\doibase 10.1103/PhysRevE.97.012909} {\bibfield  {journal} {\bibinfo
  {journal} {Phys. Rev. E}\ }\textbf {\bibinfo {volume} {97}},\ \bibinfo
  {pages} {012909} (\bibinfo {year} {2018})}\BibitemShut {NoStop}%
\bibitem [{\citenamefont {Donev}\ \emph {et~al.}(2007)\citenamefont {Donev},
  \citenamefont {Connelly}, \citenamefont {Stillinger},\ and\ \citenamefont
  {Torquato}}]{PhysRevE.75.051304}%
  \BibitemOpen
  \bibfield  {author} {\bibinfo {author} {\bibfnamefont {A.}~\bibnamefont
  {Donev}}, \bibinfo {author} {\bibfnamefont {R.}~\bibnamefont {Connelly}},
  \bibinfo {author} {\bibfnamefont {F.~H.}\ \bibnamefont {Stillinger}}, \ and\
  \bibinfo {author} {\bibfnamefont {S.}~\bibnamefont {Torquato}},\ }\href
  {\doibase 10.1103/PhysRevE.75.051304} {\bibfield  {journal} {\bibinfo
  {journal} {Phys. Rev. E}\ }\textbf {\bibinfo {volume} {75}},\ \bibinfo
  {pages} {051304} (\bibinfo {year} {2007})}\BibitemShut {NoStop}%
\bibitem [{\citenamefont {Yuan}\ \emph {et~al.}(2019)\citenamefont {Yuan},
  \citenamefont {Liu}, \citenamefont {Deng},\ and\ \citenamefont
  {Li}}]{YUAN2019186}%
  \BibitemOpen
  \bibfield  {author} {\bibinfo {author} {\bibfnamefont {Y.}~\bibnamefont
  {Yuan}}, \bibinfo {author} {\bibfnamefont {L.}~\bibnamefont {Liu}}, \bibinfo
  {author} {\bibfnamefont {W.}~\bibnamefont {Deng}}, \ and\ \bibinfo {author}
  {\bibfnamefont {S.}~\bibnamefont {Li}},\ }\href {\doibase
  https://doi.org/10.1016/j.powtec.2019.04.018} {\bibfield  {journal} {\bibinfo
   {journal} {Powder Technology}\ }\textbf {\bibinfo {volume} {351}},\ \bibinfo
  {pages} {186} (\bibinfo {year} {2019})}\BibitemShut {NoStop}%
\bibitem [{\citenamefont {Maher}\ \emph {et~al.}(2022)\citenamefont {Maher},
  \citenamefont {Stillinger},\ and\ \citenamefont
  {Torquato}}]{PhysRevMaterials.6.025603}%
  \BibitemOpen
  \bibfield  {author} {\bibinfo {author} {\bibfnamefont {C.~E.}\ \bibnamefont
  {Maher}}, \bibinfo {author} {\bibfnamefont {F.~H.}\ \bibnamefont
  {Stillinger}}, \ and\ \bibinfo {author} {\bibfnamefont {S.}~\bibnamefont
  {Torquato}},\ }\href {\doibase 10.1103/PhysRevMaterials.6.025603} {\bibfield
  {journal} {\bibinfo  {journal} {Phys. Rev. Materials}\ }\textbf {\bibinfo
  {volume} {6}},\ \bibinfo {pages} {025603} (\bibinfo {year}
  {2022})}\BibitemShut {NoStop}%
\bibitem [{\citenamefont {Blouwolff}\ and\ \citenamefont
  {Fraden}(2006)}]{Blouwolff_2006}%
  \BibitemOpen
  \bibfield  {author} {\bibinfo {author} {\bibfnamefont {J.}~\bibnamefont
  {Blouwolff}}\ and\ \bibinfo {author} {\bibfnamefont {S.}~\bibnamefont
  {Fraden}},\ }\href {\doibase 10.1209/epl/i2006-10376-1} {\bibfield  {journal}
  {\bibinfo  {journal} {Europhysics Letters ({EPL})}\ }\textbf {\bibinfo
  {volume} {76}},\ \bibinfo {pages} {1095} (\bibinfo {year}
  {2006})}\BibitemShut {NoStop}%
\bibitem [{\citenamefont {Parkhouse}\ and\ \citenamefont
  {Kelly}(1995)}]{parkhouse95}%
  \BibitemOpen
  \bibfield  {author} {\bibinfo {author} {\bibfnamefont {J.~G.}\ \bibnamefont
  {Parkhouse}}\ and\ \bibinfo {author} {\bibfnamefont {A.}~\bibnamefont
  {Kelly}},\ }\href@noop {} {\bibfield  {journal} {\bibinfo  {journal} {Proc.
  Roy. Soc. London A}\ }\textbf {\bibinfo {volume} {451}},\ \bibinfo {pages}
  {737} (\bibinfo {year} {1995})}\BibitemShut {NoStop}%
\bibitem [{\citenamefont {Philipse}(1996)}]{phi96}%
  \BibitemOpen
  \bibfield  {author} {\bibinfo {author} {\bibfnamefont {A.~P.}\ \bibnamefont
  {Philipse}},\ }\href@noop {} {\bibfield  {journal} {\bibinfo  {journal}
  {Langmuir}\ }\textbf {\bibinfo {volume} {12}},\ \bibinfo {pages} {1127}
  (\bibinfo {year} {1996})}\BibitemShut {NoStop}%
\bibitem [{\citenamefont {Freeman}\ \emph {et~al.}(2019)\citenamefont
  {Freeman}, \citenamefont {Peterson}, \citenamefont {Cao}, \citenamefont
  {Wang}, \citenamefont {Franklin},\ and\ \citenamefont
  {Weeks}}]{freeman19:_random}%
  \BibitemOpen
  \bibfield  {author} {\bibinfo {author} {\bibfnamefont {J.~O.}\ \bibnamefont
  {Freeman}}, \bibinfo {author} {\bibfnamefont {S.}~\bibnamefont {Peterson}},
  \bibinfo {author} {\bibfnamefont {C.}~\bibnamefont {Cao}}, \bibinfo {author}
  {\bibfnamefont {Y.}~\bibnamefont {Wang}}, \bibinfo {author} {\bibfnamefont
  {S.~V.}\ \bibnamefont {Franklin}}, \ and\ \bibinfo {author} {\bibfnamefont
  {E.~R.}\ \bibnamefont {Weeks}},\ }\href@noop {} {\bibfield  {journal}
  {\bibinfo  {journal} {Granular Matter}\ }\textbf {\bibinfo {volume} {21}},\
  \bibinfo {pages} {84} (\bibinfo {year} {2019})}\BibitemShut {NoStop}%
\bibitem [{\citenamefont {Williams}\ and\ \citenamefont
  {Philipse}(2003)}]{willi03}%
  \BibitemOpen
  \bibfield  {author} {\bibinfo {author} {\bibfnamefont {S.~R.}\ \bibnamefont
  {Williams}}\ and\ \bibinfo {author} {\bibfnamefont {A.~P.}\ \bibnamefont
  {Philipse}},\ }\href@noop {} {\bibfield  {journal} {\bibinfo  {journal}
  {Phys. Rev. E}\ }\textbf {\bibinfo {volume} {67}},\ \bibinfo {pages} {51301}
  (\bibinfo {year} {2003})}\BibitemShut {NoStop}%
\bibitem [{\citenamefont {Heussinger}(2020)}]{PhysRevE.102.022903}%
  \BibitemOpen
  \bibfield  {author} {\bibinfo {author} {\bibfnamefont {C.}~\bibnamefont
  {Heussinger}},\ }\href {\doibase 10.1103/PhysRevE.102.022903} {\bibfield
  {journal} {\bibinfo  {journal} {Phys. Rev. E}\ }\textbf {\bibinfo {volume}
  {102}},\ \bibinfo {pages} {022903} (\bibinfo {year} {2020})}\BibitemShut
  {NoStop}%
\bibitem [{\citenamefont {Heussinger}(2021)}]{PhysRevE.103.052903}%
  \BibitemOpen
  \bibfield  {author} {\bibinfo {author} {\bibfnamefont {C.}~\bibnamefont
  {Heussinger}},\ }\href {\doibase 10.1103/PhysRevE.103.052903} {\bibfield
  {journal} {\bibinfo  {journal} {Phys. Rev. E}\ }\textbf {\bibinfo {volume}
  {103}},\ \bibinfo {pages} {052903} (\bibinfo {year} {2021})}\BibitemShut
  {NoStop}%
\bibitem [{\citenamefont {Zhao}\ \emph {et~al.}(2012)\citenamefont {Zhao},
  \citenamefont {Li}, \citenamefont {Zou},\ and\ \citenamefont
  {Yu}}]{zhao12:_dense}%
  \BibitemOpen
  \bibfield  {author} {\bibinfo {author} {\bibfnamefont {J.}~\bibnamefont
  {Zhao}}, \bibinfo {author} {\bibfnamefont {S.}~\bibnamefont {Li}}, \bibinfo
  {author} {\bibfnamefont {R.}~\bibnamefont {Zou}}, \ and\ \bibinfo {author}
  {\bibfnamefont {A.}~\bibnamefont {Yu}},\ }\href@noop {} {\bibfield  {journal}
  {\bibinfo  {journal} {Soft Matter}\ }\textbf {\bibinfo {volume} {8}},\
  \bibinfo {pages} {1003} (\bibinfo {year} {2012})}\BibitemShut {NoStop}%
\bibitem [{\citenamefont {Tangri}\ \emph {et~al.}(2017)\citenamefont {Tangri},
  \citenamefont {Guo},\ and\ \citenamefont {Curtis}}]{tangri2017packing}%
  \BibitemOpen
  \bibfield  {author} {\bibinfo {author} {\bibfnamefont {H.}~\bibnamefont
  {Tangri}}, \bibinfo {author} {\bibfnamefont {Y.}~\bibnamefont {Guo}}, \ and\
  \bibinfo {author} {\bibfnamefont {J.~S.}\ \bibnamefont {Curtis}},\
  }\href@noop {} {\bibfield  {journal} {\bibinfo  {journal} {Powder
  Technology}\ }\textbf {\bibinfo {volume} {317}},\ \bibinfo {pages} {72}
  (\bibinfo {year} {2017})}\BibitemShut {NoStop}%
\bibitem [{\citenamefont {Gravish}\ \emph {et~al.}(2012)\citenamefont
  {Gravish}, \citenamefont {Franklin}, \citenamefont {Hu},\ and\ \citenamefont
  {Goldman}}]{PhysRevLett.108.208001}%
  \BibitemOpen
  \bibfield  {author} {\bibinfo {author} {\bibfnamefont {N.}~\bibnamefont
  {Gravish}}, \bibinfo {author} {\bibfnamefont {S.~V.}\ \bibnamefont
  {Franklin}}, \bibinfo {author} {\bibfnamefont {D.~L.}\ \bibnamefont {Hu}}, \
  and\ \bibinfo {author} {\bibfnamefont {D.~I.}\ \bibnamefont {Goldman}},\
  }\href {\doibase 10.1103/PhysRevLett.108.208001} {\bibfield  {journal}
  {\bibinfo  {journal} {Phys. Rev. Lett.}\ }\textbf {\bibinfo {volume} {108}},\
  \bibinfo {pages} {208001} (\bibinfo {year} {2012})}\BibitemShut {NoStop}%
\bibitem [{\citenamefont {Murphy}\ \emph {et~al.}(2016)\citenamefont {Murphy},
  \citenamefont {Reiser}, \citenamefont {Choksy}, \citenamefont {Singer},\ and\
  \citenamefont {Jaeger}}]{murphy16:_frees_z}%
  \BibitemOpen
  \bibfield  {author} {\bibinfo {author} {\bibfnamefont {K.~A.}\ \bibnamefont
  {Murphy}}, \bibinfo {author} {\bibfnamefont {N.}~\bibnamefont {Reiser}},
  \bibinfo {author} {\bibfnamefont {D.}~\bibnamefont {Choksy}}, \bibinfo
  {author} {\bibfnamefont {C.~E.}\ \bibnamefont {Singer}}, \ and\ \bibinfo
  {author} {\bibfnamefont {H.~M.}\ \bibnamefont {Jaeger}},\ }\href@noop {}
  {\bibfield  {journal} {\bibinfo  {journal} {Granular Matter}\ }\textbf
  {\bibinfo {volume} {18}},\ \bibinfo {pages} {26} (\bibinfo {year}
  {2016})}\BibitemShut {NoStop}%
\bibitem [{\citenamefont {{Bar\'es, Jonathan}}\ \emph
  {et~al.}(2017)\citenamefont {{Bar\'es, Jonathan}}, \citenamefont {{Zhao,
  Yuchen}}, \citenamefont {{Renouf, Mathieu}}, \citenamefont {{Dierichs,
  Karola}},\ and\ \citenamefont {{Behringer, Robert}}}]{refId0}%
  \BibitemOpen
  \bibfield  {author} {\bibinfo {author} {\bibnamefont {{Bar\'es, Jonathan}}},
  \bibinfo {author} {\bibnamefont {{Zhao, Yuchen}}}, \bibinfo {author}
  {\bibnamefont {{Renouf, Mathieu}}}, \bibinfo {author} {\bibnamefont
  {{Dierichs, Karola}}}, \ and\ \bibinfo {author} {\bibnamefont {{Behringer,
  Robert}}},\ }\href {\doibase 10.1051/epjconf/201714006021} {\bibfield
  {journal} {\bibinfo  {journal} {EPJ Web Conf.}\ }\textbf {\bibinfo {volume}
  {140}},\ \bibinfo {pages} {06021} (\bibinfo {year} {2017})}\BibitemShut
  {NoStop}%
\bibitem [{\citenamefont {Dierichs}\ and\ \citenamefont
  {Menges}(2021)}]{dierichs21:_desig}%
  \BibitemOpen
  \bibfield  {author} {\bibinfo {author} {\bibfnamefont {K.}~\bibnamefont
  {Dierichs}}\ and\ \bibinfo {author} {\bibfnamefont {A.}~\bibnamefont
  {Menges}},\ }\href@noop {} {\bibfield  {journal} {\bibinfo  {journal}
  {Bioinspir. Biomim.}\ }\textbf {\bibinfo {volume} {16}},\ \bibinfo {pages}
  {065010} (\bibinfo {year} {2021})}\BibitemShut {NoStop}%
\bibitem [{\citenamefont {Desmond}\ and\ \citenamefont
  {Franklin}(2006)}]{desmond06:_jammin}%
  \BibitemOpen
  \bibfield  {author} {\bibinfo {author} {\bibfnamefont {K.}~\bibnamefont
  {Desmond}}\ and\ \bibinfo {author} {\bibfnamefont {S.~V.}\ \bibnamefont
  {Franklin}},\ }\href@noop {} {\bibfield  {journal} {\bibinfo  {journal}
  {Phys. Rev. E}\ }\textbf {\bibinfo {volume} {73}},\ \bibinfo {pages} {031306}
  (\bibinfo {year} {2006})}\BibitemShut {NoStop}%
\bibitem [{\citenamefont {Trepanier}\ and\ \citenamefont
  {Franklin}(2010{\natexlab{a}})}]{trepanier10:_colum}%
  \BibitemOpen
  \bibfield  {author} {\bibinfo {author} {\bibfnamefont {M.}~\bibnamefont
  {Trepanier}}\ and\ \bibinfo {author} {\bibfnamefont {S.~V.}\ \bibnamefont
  {Franklin}},\ }\href@noop {} {\bibfield  {journal} {\bibinfo  {journal}
  {Phys. Rev. E}\ }\textbf {\bibinfo {volume} {82}},\ \bibinfo {pages} {011308}
  (\bibinfo {year} {2010}{\natexlab{a}})}\BibitemShut {NoStop}%
\bibitem [{\citenamefont {Sarate}\ \emph {et~al.}(2022)\citenamefont {Sarate},
  \citenamefont {Murthy},\ and\ \citenamefont {Sharma}}]{sarate22:_colum}%
  \BibitemOpen
  \bibfield  {author} {\bibinfo {author} {\bibfnamefont {P.~S.}\ \bibnamefont
  {Sarate}}, \bibinfo {author} {\bibfnamefont {T.~G.}\ \bibnamefont {Murthy}},
  \ and\ \bibinfo {author} {\bibfnamefont {P.}~\bibnamefont {Sharma}},\
  }\href@noop {} {\bibfield  {journal} {\bibinfo  {journal} {Soft Matter}\
  }\textbf {\bibinfo {volume} {18}},\ \bibinfo {pages} {2054} (\bibinfo {year}
  {2022})}\BibitemShut {NoStop}%
\bibitem [{\citenamefont {Tapia-McClung}\ and\ \citenamefont
  {Zenit}(2012)}]{PhysRevE.85.061304}%
  \BibitemOpen
  \bibfield  {author} {\bibinfo {author} {\bibfnamefont {H.}~\bibnamefont
  {Tapia-McClung}}\ and\ \bibinfo {author} {\bibfnamefont {R.}~\bibnamefont
  {Zenit}},\ }\href {\doibase 10.1103/PhysRevE.85.061304} {\bibfield  {journal}
  {\bibinfo  {journal} {Phys. Rev. E}\ }\textbf {\bibinfo {volume} {85}},\
  \bibinfo {pages} {061304} (\bibinfo {year} {2012})}\BibitemShut {NoStop}%
\bibitem [{\citenamefont {Zhao}\ \emph {et~al.}(2018)\citenamefont {Zhao},
  \citenamefont {An}, \citenamefont {Gou}, \citenamefont {Zhao},\ and\
  \citenamefont {Yang}}]{zhao18:_atten}%
  \BibitemOpen
  \bibfield  {author} {\bibinfo {author} {\bibfnamefont {H.}~\bibnamefont
  {Zhao}}, \bibinfo {author} {\bibfnamefont {X.}~\bibnamefont {An}}, \bibinfo
  {author} {\bibfnamefont {D.}~\bibnamefont {Gou}}, \bibinfo {author}
  {\bibfnamefont {B.}~\bibnamefont {Zhao}}, \ and\ \bibinfo {author}
  {\bibfnamefont {R.}~\bibnamefont {Yang}},\ }\href@noop {} {\bibfield
  {journal} {\bibinfo  {journal} {Soft Matter}\ }\textbf {\bibinfo {volume}
  {14}},\ \bibinfo {pages} {4404} (\bibinfo {year} {2018})}\BibitemShut
  {NoStop}%
\bibitem [{\citenamefont
  {Stukowski}(2010)}]{stukowski10:_visual_ovito_open_visual_tool}%
  \BibitemOpen
  \bibfield  {author} {\bibinfo {author} {\bibfnamefont {A.}~\bibnamefont
  {Stukowski}},\ }\href@noop {} {\bibfield  {journal} {\bibinfo  {journal}
  {Modelling Simul. Mater. Sci. Eng.}\ }\textbf {\bibinfo {volume} {18}},\
  \bibinfo {pages} {015012} (\bibinfo {year} {2010})}\BibitemShut {NoStop}%
\bibitem [{\citenamefont {Cundall}\ and\ \citenamefont
  {Strack}(1979)}]{cundall79}%
  \BibitemOpen
  \bibfield  {author} {\bibinfo {author} {\bibfnamefont {P.~A.}\ \bibnamefont
  {Cundall}}\ and\ \bibinfo {author} {\bibfnamefont {O.~D.~L.}\ \bibnamefont
  {Strack}},\ }\href@noop {} {\bibfield  {journal} {\bibinfo  {journal}
  {G\'eotechn.}\ }\textbf {\bibinfo {volume} {29}},\ \bibinfo {pages} {47}
  (\bibinfo {year} {1979})}\BibitemShut {NoStop}%
\bibitem [{\citenamefont {Santos}\ \emph {et~al.}(2020)\citenamefont {Santos},
  \citenamefont {Bolintineanu}, \citenamefont {Grest}, \citenamefont {Lechman},
  \citenamefont {Plimpton}, \citenamefont {Srivastava},\ and\ \citenamefont
  {Silbert}}]{PhysRevE.102.032903}%
  \BibitemOpen
  \bibfield  {author} {\bibinfo {author} {\bibfnamefont {A.~P.}\ \bibnamefont
  {Santos}}, \bibinfo {author} {\bibfnamefont {D.~S.}\ \bibnamefont
  {Bolintineanu}}, \bibinfo {author} {\bibfnamefont {G.~S.}\ \bibnamefont
  {Grest}}, \bibinfo {author} {\bibfnamefont {J.~B.}\ \bibnamefont {Lechman}},
  \bibinfo {author} {\bibfnamefont {S.~J.}\ \bibnamefont {Plimpton}}, \bibinfo
  {author} {\bibfnamefont {I.}~\bibnamefont {Srivastava}}, \ and\ \bibinfo
  {author} {\bibfnamefont {L.~E.}\ \bibnamefont {Silbert}},\ }\href {\doibase
  10.1103/PhysRevE.102.032903} {\bibfield  {journal} {\bibinfo  {journal}
  {Phys. Rev. E}\ }\textbf {\bibinfo {volume} {102}},\ \bibinfo {pages}
  {032903} (\bibinfo {year} {2020})}\BibitemShut {NoStop}%
\bibitem [{Note1()}]{Note1}%
  \BibitemOpen
  \bibinfo {note} {With the chosen boundary conditions the setting also
  resembles the dam-break problem, where collapse occurs in only one
  direction.}\BibitemShut {Stop}%
\bibitem [{Note2()}]{Note2}%
  \BibitemOpen
  \bibinfo {note} {The reason for the value $8+2f=2(4+f)$ is constraint
  counting: every rod has 4 degrees of freedom (excluding rotations around and
  translations along the long axis). The longitudinal translation is only
  counted for the $f$ rods that are longitudinally constrained via end contacts
  on both ends.}\BibitemShut {Stop}%
\bibitem [{\citenamefont {Trepanier}\ and\ \citenamefont
  {Franklin}(2010{\natexlab{b}})}]{PhysRevE.82.011308}%
  \BibitemOpen
  \bibfield  {author} {\bibinfo {author} {\bibfnamefont {M.}~\bibnamefont
  {Trepanier}}\ and\ \bibinfo {author} {\bibfnamefont {S.~V.}\ \bibnamefont
  {Franklin}},\ }\href {\doibase 10.1103/PhysRevE.82.011308} {\bibfield
  {journal} {\bibinfo  {journal} {Phys. Rev. E}\ }\textbf {\bibinfo {volume}
  {82}},\ \bibinfo {pages} {011308} (\bibinfo {year}
  {2010}{\natexlab{b}})}\BibitemShut {NoStop}%
\end{thebibliography}


\end{document}